\documentclass[aps,pra,twocolumn,superscriptaddress,nofootinbib]{revtex4-2}

\usepackage{amsmath,amssymb,amsfonts}
\usepackage{graphicx}
\usepackage{bm}
\usepackage{hyperref}
\usepackage{xcolor}

\begin{document}

\title{Collective Chirped STIRAP in a Pair of Three-Level Atoms: Dressed-Manifold Dynamics and Compensation of the Rydberg-Rydberg Interaction-Induced Detuning}

\author{Vladimir V. Malinovsky}
\affiliation{Department of Physics, Stevens Institute of Technology, Hoboken, New Jersey 07030, USA}

\author{Svetlana A. Malinovskaya}
\affiliation{Department of Physics, Stevens Institute of Technology, Hoboken, New Jersey 07030, USA}

\date{\today}

\begin{abstract}
We investigate collective chirped stimulated Raman adiabatic passage (STIRAP) in a pair of identical three-level atoms using a symmetric six-state model. Numerical simulations reveal a pronounced oscillatory dependence of the population transfer to the doubly excited Rydberg state on the peak Rabi frequency, indicating dynamics beyond the conventional single-dark-state description of STIRAP. To identify the underlying mechanism, we develop a dressed-manifold description based on gauge-invariant projections onto nearly degenerate instantaneous dressed manifolds. The analysis demonstrates that the wavefunction remains predominantly confined to a two-dimensional dark manifold while becoming transiently coupled to a neighboring bright manifold during the pulse-overlap interval. We further show that in the dressed-manifold representation, the Rydberg--Rydberg interaction modifies the collective resonance structure while preserving the dark--bright manifold dynamics responsible for the oscillatory transfer; an appropriately chosen frequency chirp compensates the interaction-induced detuning and restores efficient population transfer. 
This dressed-manifold picture provides a consistent interpretation of oscillatory collective transfer and establishes a framework for controlling adiabatic dynamics in interacting multilevel systems by linearly chirped STIRAP, opening a route toward manifold-based control of Rydberg-mediated quantum gates, correlated-state preparation, and other coherent operations in interacting atomic systems.

\end{abstract}

\maketitle

\section{Introduction}

Coherent excitation of Rydberg states is a central tool in modern atomic, molecular, and optical physics~\cite{Gallagher1994,Saffman2010,Adams2020,Browaeys2020,Morgado2021}. Because Rydberg atoms possess large transition dipole moments, long radiative lifetimes, and strong long-range interactions, they provide a versatile platform for quantum information processing, quantum simulation, nonlinear quantum optics, precision metrology, and quantum sensing.  In particular, the interaction-induced energy shift of multiply excited
Rydberg states enables blockade physics, conditional quantum gates,
entanglement generation, and controlled preparation of correlated many-body
states~\cite{Jaksch2000,Lukin2001,Urban2009,Gaetan2009,Wilk2010,
Isenhower2010,Maller2015,Levine2019,Carrasco2026}. In conditional
Rydberg gates, this interaction-induced shift can be exploited as a resource
for state-dependent phase accumulation~\cite{Carrasco2026}; in coherent
population-transfer protocols, however, the same interaction can modify the
resonance conditions required for efficient excitation of the Rydberg-pair
state. 

Stimulated Raman adiabatic passage (STIRAP) is among the most robust methods for coherent population transfer in multilevel quantum systems~\cite{Kuklinski1989,Gaubatz1990,Bergmann1998,Vitanov2001,Vitanov2017}. In a three-level ladder atom, delayed pump and Stokes pulses can transfer population from the ground state to an upper excited state while suppressing transient occupation of the lossy intermediate state. The essential mechanism is adiabatic following of a single dark state, whose composition evolves continuously from the initial state to the target state under counterintuitive pulse ordering. When the two-photon resonance condition is satisfied and the pulses vary sufficiently slowly, the transfer efficiency can approach unity and remains relatively insensitive to moderate variations of pulse amplitude and timing. Extensions of the same adiabatic framework enable robust preparation of coherent superpositions and optimized or accelerated protocols~\cite{Unanyan1998,Kis2005,Giannelli2014,Baksic2016}.

Frequency chirping provides an additional degree of control in adiabatic passage~\cite{Dridi2009,LiuMalinovskaya2015,Chathanathil2023}. By allowing the instantaneous optical detunings to vary in time, chirped STIRAP can compensate static detunings, enlarge the region of efficient transfer, and modify the structure of avoided crossings in the instantaneous dressed spectrum. In a single ladder atom, this role of chirping is often interpreted as dynamic restoration of the two-photon resonance condition. In interacting atomic systems, however, the effect of chirp is potentially richer because it modifies not only a single dark-state trajectory but an entire collective dressed-state landscape.

The simplest system in which this additional structure appears is a pair of identical three-level ladder atoms driven by the same pump and Stokes fields. Even when the Rydberg--Rydberg interaction is absent, the two-atom Hilbert space is not equivalent to two independent copies of the single-atom problem. Starting from the collective ground state, the dynamics remains in a six-dimensional symmetric subspace containing the states
\[
|11\rangle,\quad |S_{12}\rangle,\quad |22\rangle,\quad
|S_{13}\rangle,\quad |S_{23}\rangle,\quad |33\rangle ,
\]
where the symmetric states describe indistinguishable excitation of either atom. This collective manifold forms a network of coupled states rather than a simple three-state ladder, in close analogy with other collective light--matter systems organized by permutation symmetry~\cite{Dicke1954,TavisCummings1968}. Related collective Rydberg dynamics under blockade and electromagnetically induced transparency have been analyzed in Refs.~\cite{Fleischhauer2005,Muller2009,Pohl2010,RaoMolmer2014}. Consequently, the conventional single-dark-state interpretation of STIRAP is no longer sufficient.

A striking manifestation of this collective structure is the appearance of oscillatory transfer bands as the peak Rabi frequency is varied. For an isolated three-level atom, increasing the pulse amplitude beyond the adiabatic threshold generally improves population transfer monotonically. In the two-atom collective system, by contrast, the final population of the doubly excited Rydberg state exhibits pronounced maxima and minima as a function of the same control parameter. The persistence of this oscillatory structure beyond the onset of efficient transfer indicates that the dynamics is governed not simply by the degree of conventional adiabaticity, but by coherent evolution within the collective dressed-state structure.

The central result of this work is that collective two-atom STIRAP is governed not by adiabatic following of a single dark state, but by coherent evolution within and between multidimensional dressed manifolds: %Although the connectivity of the six-state Hamiltonian naturally suggests interference between alternative bare-state excitation pathways connecting $|11\rangle$ and $|33\rangle$, detailed numerical analysis shows that this interpretation is incomplete. The dynamics is  organized into
a two-dimensional dark manifold, a neighboring bright manifold, and an outer manifold that remains essentially inactive. The wavefunction evolves predominantly within the dark manifold but becomes transiently and coherently coupled to the neighboring bright manifold during the pulse-overlap interval. The extent to which the population returns to the dark manifold determines the final transfer efficiency and produces its oscillatory dependence on the peak Rabi frequency.

To formulate this mechanism independently of the arbitrary choice of instantaneous eigenvectors within nearly degenerate subspaces, we introduce gauge-invariant projectors onto the instantaneous dressed manifolds and define the corresponding wavefunction-projected components of the propagated state. This leads to the scalar nonadiabatic coupling
%\[
$\Gamma_{DB}(t)
=
\left|
\left\langle
\Psi_B(t)
\middle|
\frac{d}{dt}\Psi_D(t)
\right\rangle
\right|,$ 
%\]
which provides a generalized analogue of the conventional STIRAP mixing rate for the multidimensional-manifold problem. Unlike couplings defined between individual instantaneous dressed eigenvectors, this quantity is constructed directly from projector-defined components of the physical wavefunction and is therefore invariant under unitary rotations within each manifold. We show that $\Gamma_{DB}(t)$ is strongly localized in the pulse-overlap region, while coupling to the outer manifold remains negligible. Thus, the oscillatory collective transfer arises from localized dark--bright--dark manifold communication.  %rather than from a collapse of the dressed-state energy gap or simple interference between comparable bare-state pathways.

The second objective of this work is to determine how Rydberg--Rydberg interactions modify this collective transfer mechanism. The interaction shifts the energy of the doubly excited state $|33\rangle$ and  changes the collective Raman resonance condition. Importantly, the interaction does not destroy the dressed-manifold mechanism: the oscillatory structure persists, while the region of efficient transfer is displaced in parameter space. Frequency chirping  compensates this interaction-induced detuning by restoring the collective pair resonance during the interval of strongest pump--Stokes overlap.

For a two-photon detuning of the form
%\[
$\delta(t)=\delta_0-2\alpha(t-t_c),$ 
%\]
the resonance condition for the interacting pair state is
%\[
$2\delta(t_{\rm res})+V=0,$ 
%\]
which gives
%\[
$\alpha=\frac{2\delta_0+V}{4(t_{\rm res}-t_c)}.$ 
%\]
For zero initial two-photon detuning and counterintuitive pulses separated by $t_d$, choosing the resonance time at the center of the pulse-overlap interval yields the simple design rule
%\[
$\alpha=-\frac{V}{2t_d}.$ 
%\]
Numerical propagation of the complete six-state system confirms that this chirp restores efficient population transfer at substantially lower driving strengths while preserving the characteristic oscillatory bands.

The resulting physical picture separates two distinct aspects of collective chirped STIRAP. The dressed-manifold structure determines \emph{how} population is transported through the collective Hilbert space and is responsible for the oscillatory dependence of the transfer efficiency on the driving strength. The Rydberg--Rydberg interaction, in contrast, determines \emph{where} the collective Raman resonance occurs.  The frequency chirp provides a means of dynamically restoring that resonance without altering the underlying dressed-manifold mechanism. % This separation between coherent dressed-manifold transport and interaction-induced resonance control constitutes the principal physical picture developed in this work.

The paper is organized as follows. Section~II introduces the two-atom model, the symmetric collective basis, and the time-dependent Hamiltonian. Section~III presents the collective chirped STIRAP dynamics in the absence of Rydberg--Rydberg interactions and identifies the oscillatory transfer behavior. Section~IV develops the dressed-manifold interpretation and introduces the wavefunction-projected nonadiabatic coupling responsible for dark--bright manifold communication. Section~V analyzes the influence of the Rydberg--Rydberg interaction and derives the chirp-compensation condition. Section~VI summarizes the results and discusses their implications for quantum coherent  control of interacting multilevel systems.

%%%%%%%%%%%%%%%%%%%%%%%%%%%%%%%%%%%%%%%%%%%%%%%%%%%%%%%%%%%%%%%%%%%%%%%%%%%%%%
\section{Theoretical Model}

\subsection{Microscopic Two-Atom Hamiltonian and Symmetry Reduction}

\begin{figure*}[t]
\centering
\includegraphics[width=1.5\columnwidth]
{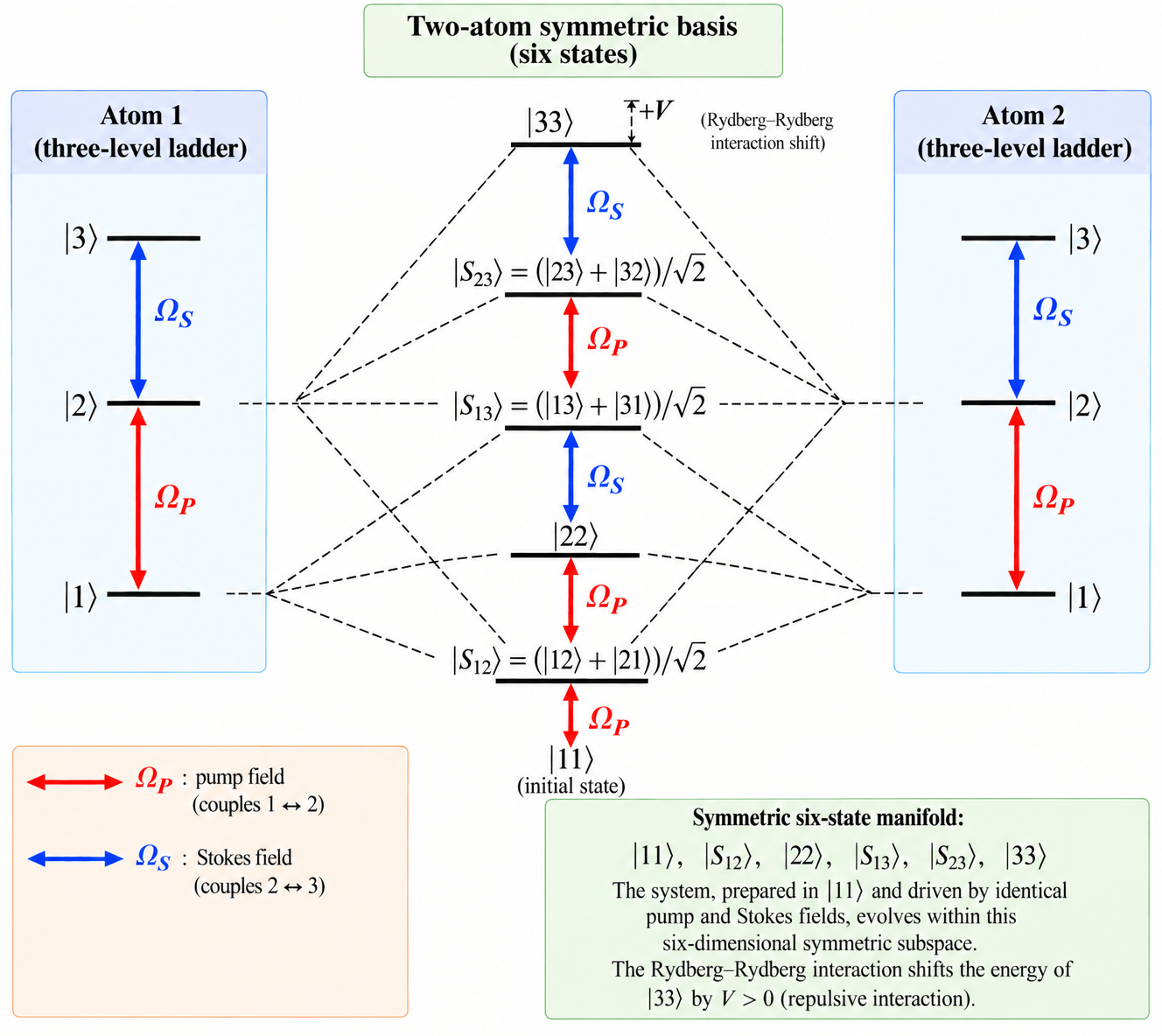}
\caption{Schematic of the collectively driven two-atom ladder system. Identical pump $\Omega_P$, (red) and Stokes $\Omega_S$, (blue) fields couple the six symmetric two-atom states. The doubly excited Rydberg state $|33\rangle$ is shifted by the repulsive interaction $V>0$.
}
\label{fig:scheme}
\end{figure*}

We consider two identical ladder-type three-level atoms interacting with
time-dependent pump and Stokes laser fields under the electric-dipole and
rotating-wave approximations~\cite{Bergmann1998,Vitanov2017,Shore2011}, see Fig.(\ref{fig:scheme}).

Each atom possesses a ground state
$|1\rangle$, an intermediate state $|2\rangle$, and a Rydberg state
$|3\rangle$. The pump laser couples the transition
$|1\rangle\leftrightarrow|2\rangle$, while the Stokes laser couples
$|2\rangle\leftrightarrow|3\rangle$.

For a single atom, the Hamiltonian may be written as

\begin{equation}
H^{(1)}(t)
=
H_0
+
H_{\mathrm{int}}(t),
\label{eq:Hsingle}
\end{equation}
where $H_0$ contains the atomic energies and
$H_{\mathrm{int}}(t)$ describes the interaction with the two laser fields.
Within the rotating-wave approximation, the interaction Hamiltonian takes the
form

\begin{equation}
H_{\mathrm{int}}(t)
=
\frac{\hbar}{2}
\left(
\begin{array}{ccc}
0
&
\Omega_P(t)
&
0
\\
\Omega_P(t)
&
2\Delta(t)
&
\Omega_S(t)
\\
0
&
\Omega_S(t)
&
2\Delta(t)+2\delta(t)
\end{array}
\right),
\label{eq:HsingleMatrix}
\end{equation}
where $\Omega_P(t)$ and $\Omega_S(t)$ denote the instantaneous pump and
Stokes Rabi frequencies, respectively, while $\Delta(t)$ and
$\delta(t)$ represent the one-photon and two-photon detunings.

The Hilbert space of two noninteracting atoms is the tensor product
\begin{equation}
\mathcal H
=
\mathcal H_1
\otimes
\mathcal H_2,
\end{equation}
which has dimension %\begin{equation}
$3\times3=9.$ 
%\end{equation}
The corresponding Hamiltonian is
\begin{equation}
H_0^{(2)}
=
H^{(1)}
\otimes I
+
I\otimes H^{(1)},
\label{eq:Htwo}
\end{equation}
where the first and second terms describe the interaction of the laser fields
with atoms 1 and 2, respectively.

The Rydberg--Rydberg interaction introduces an additional two-body
contribution~\cite{Jaksch2000,Lukin2001,Saffman2010},
\begin{equation}
H_{\mathrm{RR}}
=
V_{\mathrm{RR}}
|33\rangle\langle33|,
\label{eq:HRR}
\end{equation}
where $V_{\mathrm{RR}}$ denotes the interaction energy of the doubly excited
Rydberg pair state. Unlike a single-particle energy shift, this term acts
only when both atoms occupy the Rydberg state simultaneously.

The complete microscopic Hamiltonian therefore becomes
\begin{equation}
H
=
H^{(1)}
\otimes I
+
I\otimes H^{(1)}
+
H_{\mathrm{RR}}.
\label{eq:Htotal}
\end{equation}
%%%%%%%%%%%%%%%%%%%%%%%%%%%%%%%%%%%%%%%%%%%%%%%%%%%%%%%%%%%%%%%%%%%%%%%%%%%
The natural product basis consists of the nine states
\begin{equation}
\{
|11\rangle,
|12\rangle,
|13\rangle,
|21\rangle,
|22\rangle,
|23\rangle,
|31\rangle,
|32\rangle,
|33\rangle
\}.
\end{equation}
Since the two atoms are identical and are driven by the same laser fields,
the Hamiltonian is invariant under exchange of the two atoms~\cite{Dicke1954,TavisCummings1968},
\begin{equation}
[H,\hat P_{12}]=0,
\end{equation}
where $\hat P_{12}$ is the particle-exchange operator.

Consequently, the Hilbert space separates into symmetric and antisymmetric
subspaces,
\begin{equation}
\mathcal H
=
\mathcal H_S
\oplus
\mathcal H_A,
\end{equation}
which evolve independently. 

Because the two atoms are identical and experience the same driving fields,
their exchange cannot change the Hamiltonian or the resulting dynamics.
The symmetric and antisymmetric subspaces therefore constitute distinct
dynamical sectors that are not coupled by the evolution. In particular,
starting from the symmetric ground state $|11\rangle$, the system can access
only states that are symmetric under exchange of the two atoms. For example,
the singly excited state coupled to $|11\rangle$ is
\[
|S_{12}\rangle=\frac{|12\rangle+|21\rangle}{\sqrt{2}},
\]
whereas the antisymmetric combination
\[
|A_{12}\rangle=\frac{|12\rangle-|21\rangle}{\sqrt{2}}
\]
is not coupled to the symmetric initial state. Thus, exchange symmetry
provides the physical reason why the dynamics remains confined to the
symmetric sector. Assuming the system is initially prepared in the symmetric ground state
\begin{equation}
|\Psi(0)\rangle
=
|11\rangle,
\label{ground_state}
\end{equation}
the antisymmetric sector is never populated because it is not coupled by the
Hamiltonian. The dynamics is therefore completely confined to the symmetric
six-dimensional manifold
\begin{equation}
\mathcal H_S
=
\{
|11\rangle,
|S_{12}\rangle,
|22\rangle,
|S_{13}\rangle,
|S_{23}\rangle,
|33\rangle
\},
\label{eq:symmetricbasis}
\end{equation}
where
\begin{align}
|S_{12}\rangle
&=
\frac{|12\rangle+|21\rangle}{\sqrt2},\nonumber
\\
|S_{13}\rangle
&=
\frac{|13\rangle+|31\rangle}{\sqrt2},\nonumber
\\
|S_{23}\rangle
&=
\frac{|23\rangle+|32\rangle}{\sqrt2}.
\end{align}
This symmetry reduction %decreases the dimension of the dynamical problem from nine to six while preserving all physically accessible states for symmetric excitation. More importantly, it 
reveals that the collective dynamics is not
equivalent to two independent three-level systems but instead forms a
connected six-state network whose topology is fundamentally different from the
single-atom ladder configuration. As shown in the following sections, this
collective structure is responsible for the emergence of oscillatory
population transfer and motivates the dressed-manifold description developed
later in the paper.

The effective Hamiltonian in the symmetric collective basis is presented in the
following subsection.

%%%%%%%%%%%%%%%%%%%%%%%%%%%%%%%%%%%%%%%%%%%%%%%%%%%%%%%%%%%%%%%%%%%%%%%%%%%%%%
\subsection{Effective Collective Hamiltonian}

Projecting the microscopic Hamiltonian (\ref{eq:Htotal}) onto the symmetric
subspace (\ref{eq:symmetricbasis}) yields an effective six-dimensional
Hamiltonian governing the complete collective dynamics. Writing the
wavefunction as
\begin{eqnarray}
&|\Psi(t)\rangle
=
a_{11}|11\rangle
+a_{12}|S_{12}\rangle
+a_{22}|22\rangle
+a_{13}|S_{13}\rangle+ \nonumber \\
&a_{23}|S_{23}\rangle
+a_{33}|33\rangle ,
\label{eq:wavefunction}
\end{eqnarray}
the time-dependent Schrödinger equation,
\begin{equation}
i\hbar
\frac{\partial}{\partial t}
|\Psi(t)\rangle
=
H_{\rm eff}(t)
|\Psi(t)\rangle ,
\label{eq:TDSE}
\end{equation}
is governed by the effective Hamiltonian 
\begin{equation}
\begin{split}
&H_{\rm eff}(t)
=
\frac{\hbar}{2}\times\\
&\left(
\begin{array}{cccccc}
0
&
\sqrt2\,\Omega_P
&
0
&
0
&
0
&
0
\\
\sqrt2\,\Omega_P
&
2\Delta
&
\sqrt2\,\Omega_P
&
\Omega_S
&
0
&
0
\\
0
&
\sqrt2\,\Omega_P
&
4\Delta
&
0
&
\sqrt2\,\Omega_S
&
0
\\
0
&
\Omega_S
&
0
&
2(\Delta+\delta)
&
\Omega_P
&
0
\\
0
&
0
&
\sqrt2\,\Omega_S
&
\Omega_P
&
2(2\Delta+\delta)
&
\sqrt2\,\Omega_S
\\
0
&
0
&
0
&
0
&
\sqrt2\,\Omega_S
&
4\Delta+4\delta+2V_{\rm RR}
\end{array}
\right),
\label{eq:Heff}
\end{split}
\end{equation}
where the explicit time dependence of
$\Omega_{P,S}(t)$,
$\Delta(t)$,
and
$\delta(t)$
has been omitted for compactness. 
Their definitions are provided in the next subsection.

%%%%%%%%%%%%%%%%%%%%%%%%%%%%%%%%%%%%%%%%%%%%%%%%%%%%%%%%%%%%%%%%%%%%%%%%%%%

Several features of Eq.~(\ref{eq:Heff}) require attention. First, the factors of $\sqrt2$ originate entirely from the normalization of
the symmetric basis states. Whenever a transition connects a product state
with a symmetric superposition, two indistinguishable excitation pathways add
constructively, increasing the effective coupling by a factor of $\sqrt2$.
Transitions between two symmetric states contain only a single collective
pathway and therefore retain the original single-atom Rabi frequency.

Second, the diagonal elements are simply the sums of the corresponding
single-atom detunings. For example, the state $|22\rangle$ contains two atoms
in the intermediate level and therefore acquires twice the one-photon
detuning, while the state $|S_{23}\rangle$ contains one atom in level
$|2\rangle$ and one in level $|3\rangle$, giving the combined contribution
$2(2\Delta+\delta)$ inside the matrix.

Finally, the doubly excited Rydberg state $|33\rangle$ acquires the additional
interaction energy $V_{\rm RR}$. Since this interaction is a genuine two-body
effect, it contributes only to the pair state and leaves all remaining basis
states unchanged.

%%%%%%%%%%%%%%%%%%%%%%%%%%%%%%%%%%%%%%%%%%%%%%%%%%%%%%%%%%%%%%%%%%%%%%%%%%%

Equation (\ref{eq:Heff}) immediately reveals that the collective dynamics is
topologically different from that of an isolated three-level atom. Instead of
a single ladder, the system forms a connected six-state excitation network
containing multiple collective pathways between the initial state
$|11\rangle$ and the target state $|33\rangle$. %Although this connectivity suggests the possibility of pathway interference, 
The dressed-state
analysis presented in Sec.~IV demonstrates that the dominant physical
mechanism relies on coupled instantaneous
dressed manifolds rather than independent bare-state trajectories.

%%%%%%%%%%%%%%%%%%%%%%%%%%%%%%%%%%%%%%%%%%%%%%%%%%%%%%%%%%%%%%%%%%%%%%%%%%%

For numerical convenience we introduce dimensionless time
\begin{equation}
t
\rightarrow
\frac{t}{\tau}, 
\end{equation}
where $\tau$ denotes the Gaussian pulse duration.
All frequencies are expressed in units of $\tau^{-1}$, so that the Rabi
frequencies, detunings, chirp rate, and Rydberg interaction are measured in
the same dimensionless units throughout this work.
%%%%%%%%%%%%%%%%%%%%%%%%%%%%%%%%%%%%%%%%%%%%%%%%%%%%%%%%%%%%%%%%%%%%%%%%%%%%%%
\subsection{Chirped Pulse Sequence and Numerical Propagation}

The collective dynamics is driven by a pair of delayed Gaussian laser pulses
coupling the lower and upper transitions of the ladder system. The pump field,
which couples the transition
$|1\rangle\leftrightarrow|2\rangle$,
is described by the Rabi frequency in the form
\begin{equation}
\Omega_P(t)
=
\Omega_0
\exp
\left[
-
\frac{(t-t_P)^2}{\tau^2}
\right],
\label{eq:pump}
\end{equation}
where $\Omega_0$ denotes the peak Rabi frequency,
$\tau$ is the pulse duration,
and $t_P$ specifies the temporal position of the pump pulse.

The Stokes pulse couples the transition
$|2\rangle\leftrightarrow|3\rangle$
and is described by the Rabi frequency taken as
\begin{equation}
\Omega_S(t)
=
\Omega_0
\exp
\left[
-
\frac{(t-t_S)^2}{\tau^2}
\right],
\label{eq:stokes}
\end{equation}
where
\begin{equation}
t_S=t_P-t_d,
\end{equation}
and $t_d$ is the time delay between the pulses.

Throughout this work we employ the counterintuitive pulse ordering required
for STIRAP, so that the Stokes pulse precedes the pump pulse.
Unless otherwise stated,
\begin{equation}
t_d=0.75\tau.
\end{equation}
Consequently, the maximum overlap of the two Gaussian pulses occurs at
\begin{equation}
t_{\mathrm{ov}}
=
\frac{t_P+t_S}{2}
=
t_P-\frac{t_d}{2}.
\label{eq:overlap}
\end{equation}
This overlap region plays a central role in the collective dynamics discussed
later because it is where the strongest communication between neighboring
dressed manifolds is observed.

%%%%%%%%%%%%%%%%%%%%%%%%%%%%%%%%%%%%%%%%%%%%%%%%%%%%%%%%%%%%%%%%%%%%%%%%%%%

To provide additional control over the adiabatic evolution, both laser fields
are chosen to have identical linear frequency chirps~\cite{LiuMalinovskaya2015,Chathanathil2023}. The one-photon and
two-photon detunings are therefore written as
\begin{equation}
\Delta(t)
=
\Delta_0
-
\alpha
(t-t_P),
\label{eq:Delta}
\end{equation}
and
\begin{equation}
\delta(t)
=
\delta_0
-
2\alpha
(t-t_P),
\label{eq:delta}
\end{equation}
where $\alpha$ denotes the chirp rate.
%The factor of two in the two-photon detuning reflects the accumulation of the frequency sweep along the two-photon excitation pathway.

When the Rydberg--Rydberg interaction is absent,
the resonance condition is simply
\begin{equation}
\delta(t)=0.
\end{equation}
When interactions are included, however, the doubly excited Rydberg state is
shifted by the interaction energy.
The corresponding resonance condition becomes
\begin{equation}
2\delta(t)+V_{\rm RR}=0,
\label{eq:pairres}
\end{equation}
which forms the basis of the analytical chirp-compensation condition developed
in Sec.~V.
%%%%%%%%%%%%%%%%%%%%%%%%%%%%%%%%%%%%%%%%%%%%%%%%%%%%%%%%%%%%%%%%%%%%%%%%%%%

Initially the system occupies the collective ground state, Eq.(\ref{ground_state}), corresponding to
\begin{equation}
a_{11}(0)=1,
\qquad
a_i(0)=0
\quad
(i\neq11).
\end{equation}

The time evolution is obtained by solving the coupled Schrödinger equations
\begin{equation}
i\hbar
\frac{d}{dt}
\mathbf a(t)
=
H_{\rm eff}(t)
\mathbf a(t),
\label{eq:ode}
\end{equation}
where
\[
\mathbf a(t)
=
(a_{11},
a_{12},
a_{22},
a_{13},
a_{23},
a_{33})^T .
\]
The equations are propagated numerically using a fourth-order Runge--Kutta
algorithm with sufficiently small time steps to ensure convergence.
Throughout every calculation, the normalization condition
\begin{equation}
\sum_{i=1}^{6}
|a_i(t)|^2
=
1
\label{eq:norm}
\end{equation}
is monitored continuously.
The numerical error in the total probability remains below machine precision
over the entire propagation interval.
%%%%%%%%%%%%%%%%%%%%%%%%%%%%%%%%%%%%%%%%%%%%%%%%%%%%%%%%%%%%%%%%%%%%%%%%%%%

The principal observables investigated in this work are the bare-state
populations 
%\begin{equation}
%$P_i(t)
%=
$|a_{i}(t)|^2,$ 
%\end{equation}
particulalry, the population transferred to the doubly excited Rydberg state, 
%\begin{equation}
$P_{33}(t)
=
|a_{33}(t)|^2,$ 
%\end{equation}
and, in later sections, the corresponding dressed-manifold populations
obtained by projection onto the instantaneous eigenstates of the effective
Hamiltonian.

Unless otherwise stated, all numerical simulations presented below employ
dimensionless units with $\tau=1$, the pulse centers
$t_P=t_c$,
$t_S=t_c-t_d$,
and equal peak Rabi frequencies for the pump and Stokes pulses.

%%%%%%%%%%%%%%%%%%%%%%%%%%%%%%%%%%%%%%%%%%%%%%%%%%%%%%%%%%%%%%%%%%%%%%%%%%%%%%
\section{Collective Chirped STIRAP Without Rydberg--Rydberg Interaction}

We first investigate the collective dynamics in the absence of
Rydberg--Rydberg interactions,
%
%\begin{equation}
$V_{\rm RR}=0,$ 
%\end{equation}
%
so that the oscillatory transfer mechanism can be identified independently of
interaction-induced level shifts. Throughout this section the dynamics is
governed entirely by the effective Hamiltonian introduced in Sec.~II, while
the pulse sequence follows the counterintuitive STIRAP ordering with the
Stokes pulse preceding the pump pulse.

%%%%%%%%%%%%%%%%%%%%%%%%%%%%%%%%%%%%%%%%%%%%%%%%%%%%%%%%%%%%%%%%%%%%%%%%%%%%%%
\subsection{Bare-State Population Dynamics}

The system is initially prepared in the collective ground state
$ |\Psi(0)\rangle=|11\rangle ,$ 
%\]
and evolves under the action of delayed pump and Stokes pulses.

\begin{figure*}[t]
\centering
\includegraphics[width=2.\columnwidth]%{Population_Dynamics_9Panels.pdf}
{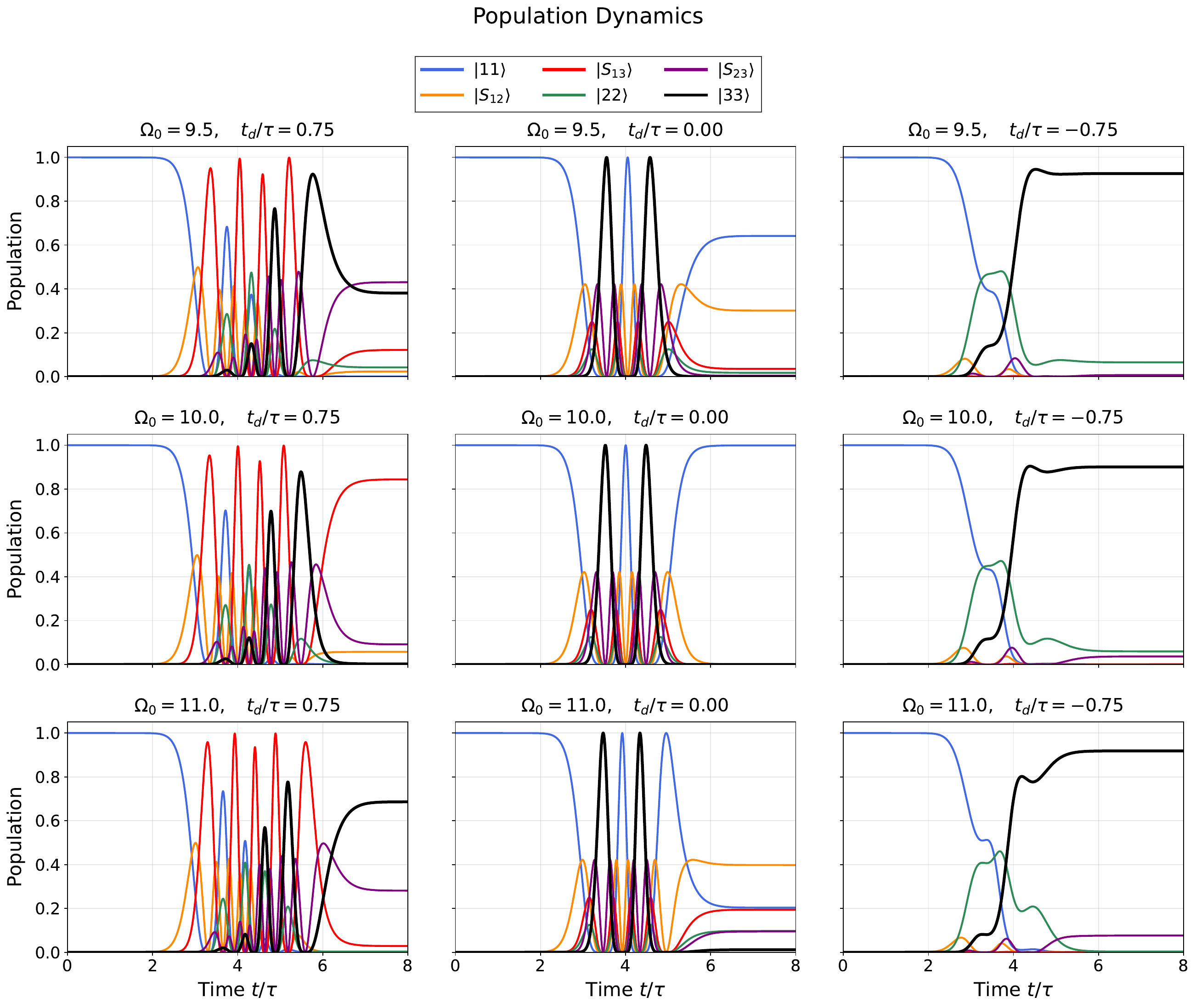}
\caption{
Time evolution of the bare-state populations for the collective
six-state system in the absence of Rydberg--Rydberg interaction: (left) intuitive sequence, (center) complete overlap, (right) counter-intuitive sequence - STIRAP. In the latter, 
the transient population of
$|22\rangle$
remains significantly smaller than that of
$|S_{13}\rangle$,
indicating that the oscillatory transfer cannot be interpreted solely as
interference between two equally populated bare-state pathways.
}

\label{fig:bare}

\end{figure*}

Figure~\ref{fig:bare} presents the time evolution of the six bare
collective-state populations for representative pulse parameters and three cases: the intuitive sequence of the pump and Stokes pulses, their simultaneous implementation with complete overlap, and counter-intuitive sequence - STIRAP. %, which demonstrate the adiabatic passage to the final, $|33\rangle$, state. 
In the latter case, as the Stokes pulse arrives before the pump pulse, population is
gradually transferred from $|11\rangle$ through the collective manifold toward the doubly excited Rydberg state $|33\rangle$. The dominant transient population develops through the intermediate state
$|13\rangle$, whereas the symmetric state
$|S_{12}\rangle$
remains only weakly populated throughout the evolution.
Likewise, the occupation of
$|S_{23}\rangle$
remains comparatively small.

These observations already suggest that the collective transfer cannot be
understood simply as interference between two equally important excitation
paths connecting
$|11\rangle$
and
$|33\rangle$.
Instead, one collective pathway appears to dominate the bare-state dynamics.

At the end of the pulse sequence the population is transferred almost entirely
to the doubly excited Rydberg state, 
%\[
$P_{33}(t_f),$ %\approx1,$ 
%\]
demonstrating efficient collective chirped STIRAP.

%%%%%%%%%%%%%%%%%%%%%%%%%%%%%%%%%%%%%%%%%%%%%%%%%%%%%%%%%%%%%%%%%%%%%%%%%%%%%%

%%%%%%%%%%%%%%%%%%%%%%%%%%%%%%%%%%%%%%%%%%%%%%%%%%%%%%%%%%%%%%%%%%%%%%%%%%%%%%
\subsection{Oscillatory Dependence on Pulse Area}

The efficiency of collective chirped STIRAP depends strongly on the peak Rabi
frequency.

To investigate this behavior systematically, the Schrödinger equation was
solved for a wide range of pulse amplitudes while keeping all remaining
parameters fixed.

\begin{figure}[t]

\centering

\includegraphics[width=\columnwidth]{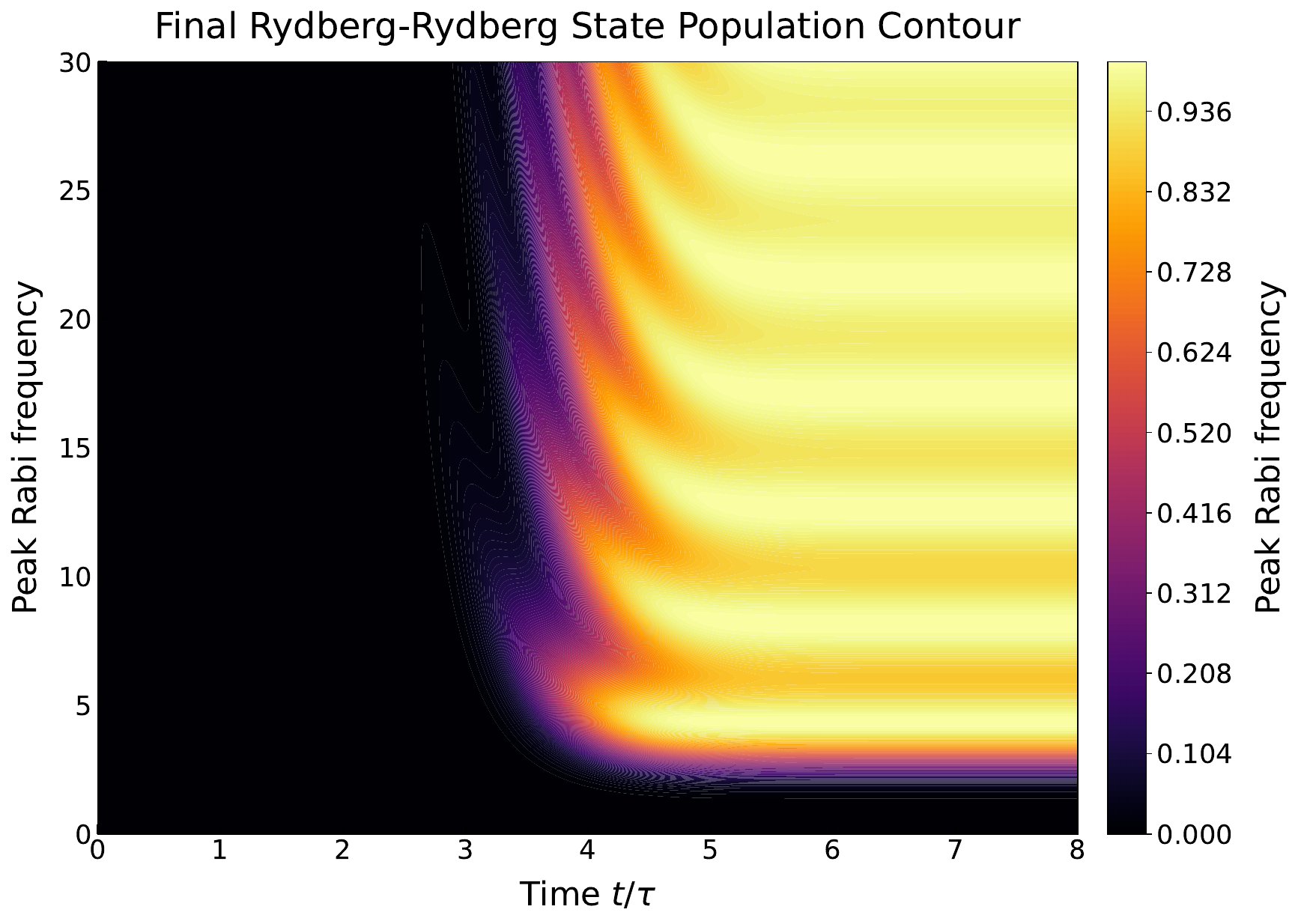}

\caption{
Contour plot of the Rydberg-Rydberg state population
$P_{33}(t)$
as a function of propagation time and peak Rabi frequency.
The alternating bright and dark bands demonstrate pronounced oscillatory
dependence of the collective transfer on the pulse area.
}

\label{fig:Contour}

\end{figure}

Figure~\ref{fig:Contour}
shows the resulting contour plot of the population of the doubly excited
Rydberg state, 
$P_{33}(t),$ 
as a function of propagation time and the peak Rabi frequency.

Unlike conventional single-atom STIRAP, where increasing the pulse amplitude
generally improves the transfer monotonically once the adiabatic condition is
satisfied, the collective system exhibits a sequence of alternating maxima and
minima.
These oscillatory bands persist over a broad range of peak Rabi frequencies and are
clearly visible in the final transferred population.

%The corresponding final population,
%
%%P_{33}(t_f),
%\]
%is shown in Fig.~\ref{fig:FinalPopulation}.
%Instead of approaching unity monotonically, the transfer efficiency displays pronounced oscillations as the pulse area increases.

This behavior demonstrates that the collective dynamics contains an additional
coherent mechanism beyond the conventional adiabatic following of a single
dark state.
Although the connectivity of the six-state network naturally suggests
interference between different excitation pathways, the bare-state populations
presented in Fig.(\ref{fig:bare})(right) do not support a simple pathway-interference picture because
one intermediate branch remains only weakly populated during the evolution.

The physical origin of the oscillatory transfer therefore cannot be inferred
from the bare-state representation alone.
A more appropriate description requires transformation to the instantaneous
dressed-state basis, where the collective dynamics can be analyzed in terms of
coupled dressed manifolds.

%%%%%%%%%%%%%%%%%%%%%%%%%%%%%%%%%%%%%%%%%%%%%%%%%%%%%%%%%%%%%%%%%%%%%%%%%%%%%%

%%%%%%%%%%%%%%%%%%%%%%%%%%%%%%%%%%%%%%%%%%%%%%%%%%%%%%%%%%%%%%%%%%%%%%%%%%%%%%

%\begin{figure}[t]

%\centering

%\includegraphics[width=0.9\columnwidth]{Fig3_FinalPopulation.pdf}

%\caption{Final population transferred to the doubly excited Rydberg state as a function of the peak Rabi frequency. The oscillatory dependence motivates the dressed-manifold analysis developed in the following section.}

%\label{fig:FinalPopulation}

%\end{figure}

%%%%%%%%%%%%%%%%%%%%%%%%%%%%%%%%%%%%%%%%%%%%%%%%%%%%%%%%%%%%%%%%%%%%%%%%%%%%%%
\section{Dressed-Manifold Interpretation of Collective Chirped STIRAP}

\subsection{Why the Bare-State Picture Is Insufficient}

The oscillatory dependence of the collective transfer efficiency on the peak
Rabi frequency, demonstrated in Sec.~III, has no counterpart in conventional
single-atom STIRAP. % In an isolated three-level ladder system, increasing the pulse area beyond the adiabatic threshold generally improves the transfer monotonically. The alternating maxima and minima observed in the collective two-atom system therefore indicate the presence of an additional coherent mechanism.

A natural first interpretation is to attribute these oscillations to quantum
interference between different bare-state excitation pathways connecting the
initial state $|11\rangle$ with the doubly excited Rydberg state $|33\rangle$.
Inspection of the collective Hamiltonian indeed suggests two qualitatively
different routes through the six-state network,
\begin{equation}
|11\rangle
\rightarrow
|S_{12}\rangle
\rightarrow
|22\rangle
\rightarrow
|S_{23}\rangle
\rightarrow
|33\rangle ,
\label{eq:path1}
\end{equation}
and
\begin{equation}
|11\rangle
\rightarrow
|S_{12}\rangle
\rightarrow
|S_{13}\rangle
\rightarrow
|S_{23}\rangle
\rightarrow
|33\rangle .
\label{eq:path2}
\end{equation}
If both pathways carried comparable probability amplitudes, constructive and
destructive interference between them could naturally produce oscillatory
dependence of the final transfer efficiency on the pulse area.

\begin{figure}[t]
\centering
\includegraphics[width=\columnwidth]{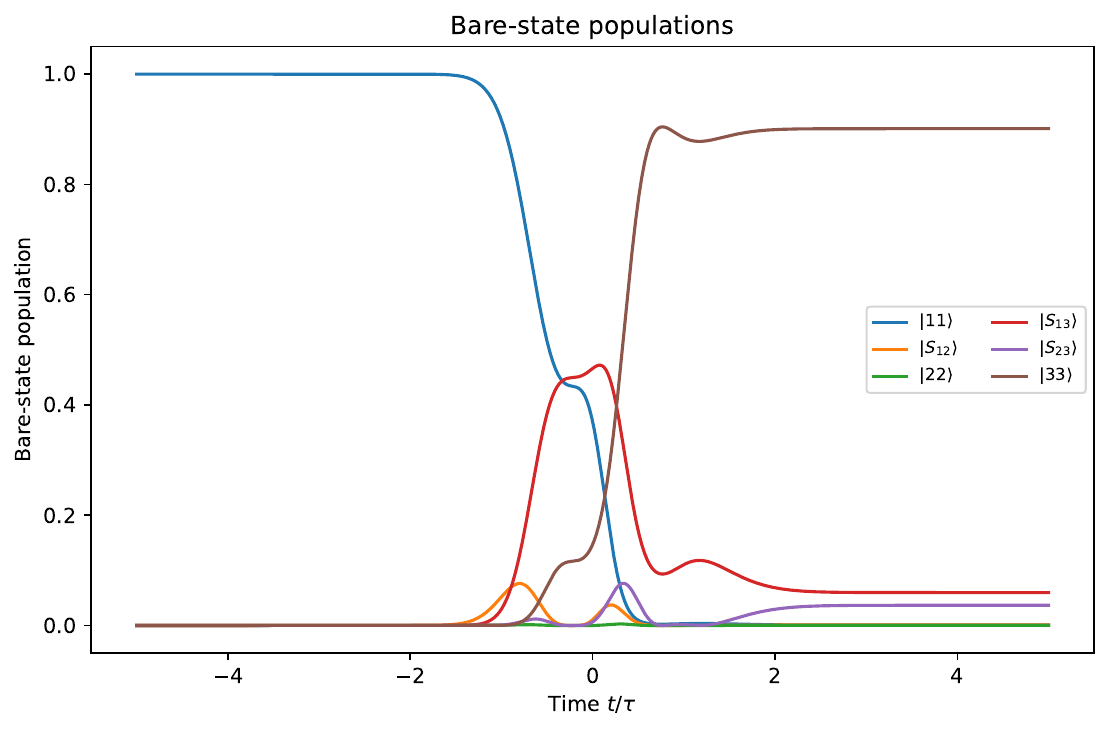}
\caption{
Time evolution of the six bare-state populations.
The dominant transient occupation occurs through the state
$|S_{13}\rangle$,
while the population of
$|22\rangle$
remains comparatively small.
}
\label{fig:bare_populations}
\end{figure}

To examine this possibility, Fig.~\ref{fig:bare_populations} shows the time
evolution of the bare-state populations for representative pulse parameters.
The results immediately reveal a pronounced asymmetry between the two
collective routes. The transient occupation of the state $|S_{13}\rangle$
becomes substantial during the pulse-overlap interval, whereas the population
of the symmetric state $|22\rangle$ remains comparatively small
throughout the evolution.

This observation has an important consequence.
Although the collective Hamiltonian permits multiple excitation pathways,
their contributions are far from equivalent.
The oscillatory transfer therefore cannot be explained simply as interference
between two bare-state trajectories possessing similar amplitudes.
The bare-state representation provides a useful description of population
flow, but it does not identify the coherent mechanism  responsible for the
oscillatory dynamics.
%The bare-state populations provide a useful description of how population is distributed among the atomic basis states, but they do not by themselves reveal the dynamical structure responsible for the collective transfer. 

During the interaction with the pump and Stokes pulses, the atomic states are continuously coupled by the applied fields, and the relevant dynamical states are therefore not the uncoupled bare atomic states, but the instantaneous field-dressed states of the time-dependent Hamiltonian. These dressed states incorporate the atom--field coupling at each instant of time and provide the natural basis for following the evolution during the pulse sequence. We therefore turn to a dressed-state analysis by diagonalizing the instantaneous Hamiltonian \(H(t)\) and examining the evolution in its adiabatic eigenstates ~\cite{BornFock1928,Kato1950,Bergmann1998,Vitanov2017}. 
%From this perspective, the failure of the bare-state populations to provide a simple interpretation of the collective transfer indicates that the relevant dynamical degrees of freedom are not the individual bare-state populations, but the dressed states and, more generally, the nearly degenerate dressed manifolds through which the evolution proceeds.

%The failure of the bare-state picture suggests that the relevant degrees of freedom are not the instantaneous populations themselves but rather the adiabatic eigenstates of the time-dependent Hamiltonian~\cite{BornFock1928,Kato1950,Bergmann1998,Vitanov2017}.
We diagonalize the instantaneous Hamiltonian
(\ref{eq:Heff})
at every propagation time,
\begin{equation}
H_{\rm eff}(t)
|\Phi_n(t)\rangle
=
E_n(t)
|\Phi_n(t)\rangle,
\qquad
n=1,\ldots,6,
\label{eq:eigenproblem}
\end{equation}
obtaining six instantaneous dressed states
$\{|\Phi_n(t)\rangle\}$ having energies   $\{{E_n(t)}\}$.

\begin{figure}[t]
\centering
\includegraphics[width=\columnwidth]{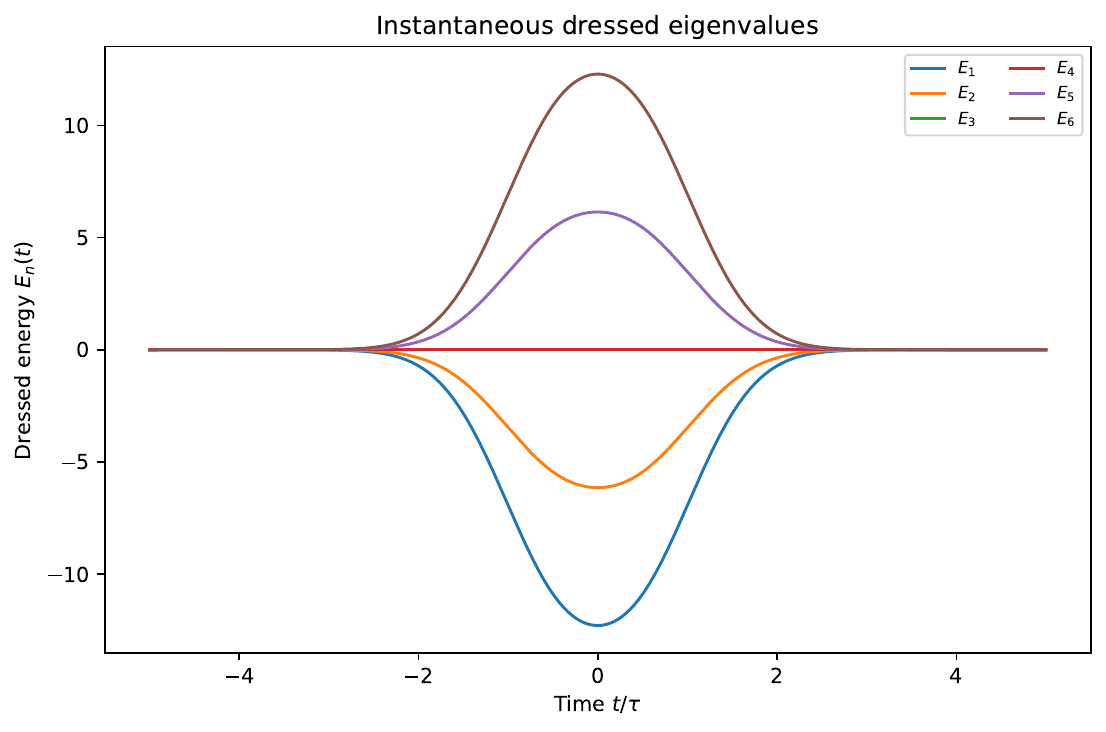}
\caption{
Instantaneous dressed-state energies obtained from the numerical
propagation.
}
\label{fig:dressed_energies}
\end{figure}

\begin{figure}[t]
\centering
\includegraphics[width=\columnwidth]{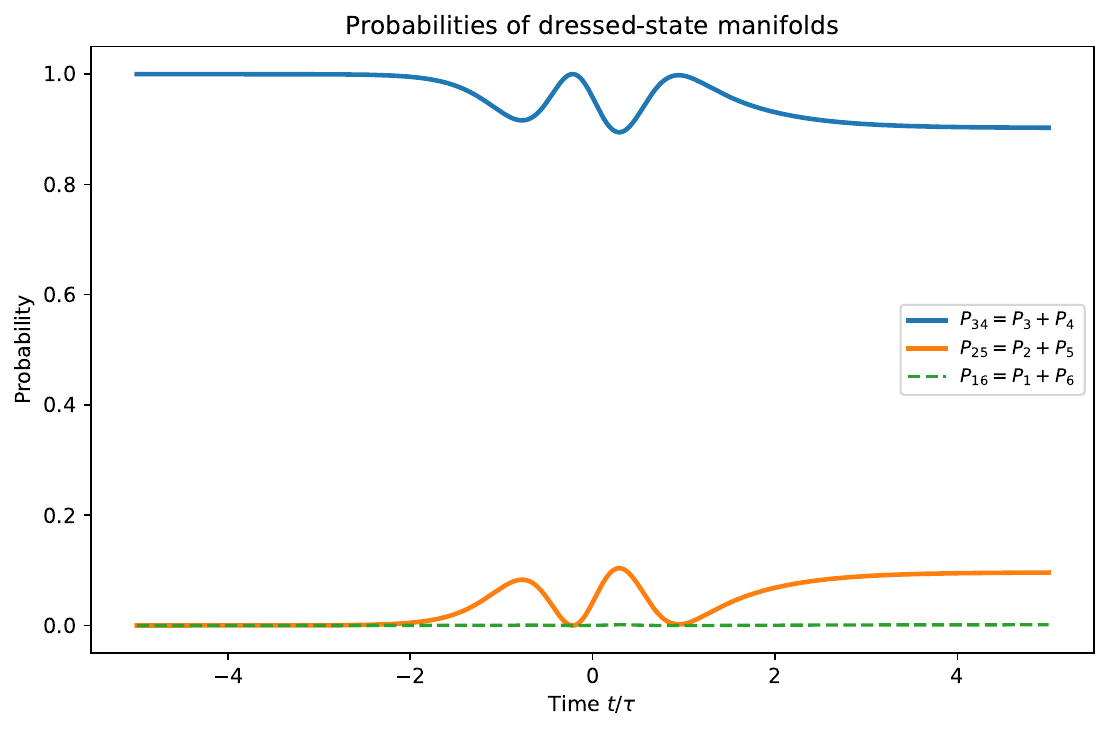}
\caption{
Instantaneous dressed-state populations obtained from the numerical
propagation.
The dressed states form three nearly degenerate pairs that evolve
collectively rather than independently.
}
\label{fig:dressed_populations}
\end{figure}

Figure~\ref{fig:dressed_energies}
shows the energies, and Figure~\ref{fig:dressed_populations}
shows the corresponding dressed-state populations,
\begin{equation}
P_n(t)
=
|\langle\Phi_n(t)|\Psi(t)\rangle|^2.
\end{equation}

A remarkable feature immediately emerges.
Rather than occupying all six dressed states,
the wavefunction is almost entirely confined to four of them.
Moreover, the dressed states naturally organize into degenerate pairs,
\[
(\Phi_1,\Phi_6),
\qquad
(\Phi_2,\Phi_5),
\qquad
(\Phi_3,\Phi_4),
\]
whose populations exhibit qualitatively different behavior.

The outer pair
$(\Phi_1,\Phi_6)$
remains negligibly populated,
whereas the remaining four dressed states carry essentially the entire
wavefunction throughout the evolution.
This behavior already indicates that the physically relevant dynamical states 
are not individual eigenstates but collective dressed manifolds.

\begin{figure}[t]
\centering
\includegraphics[width=\columnwidth]{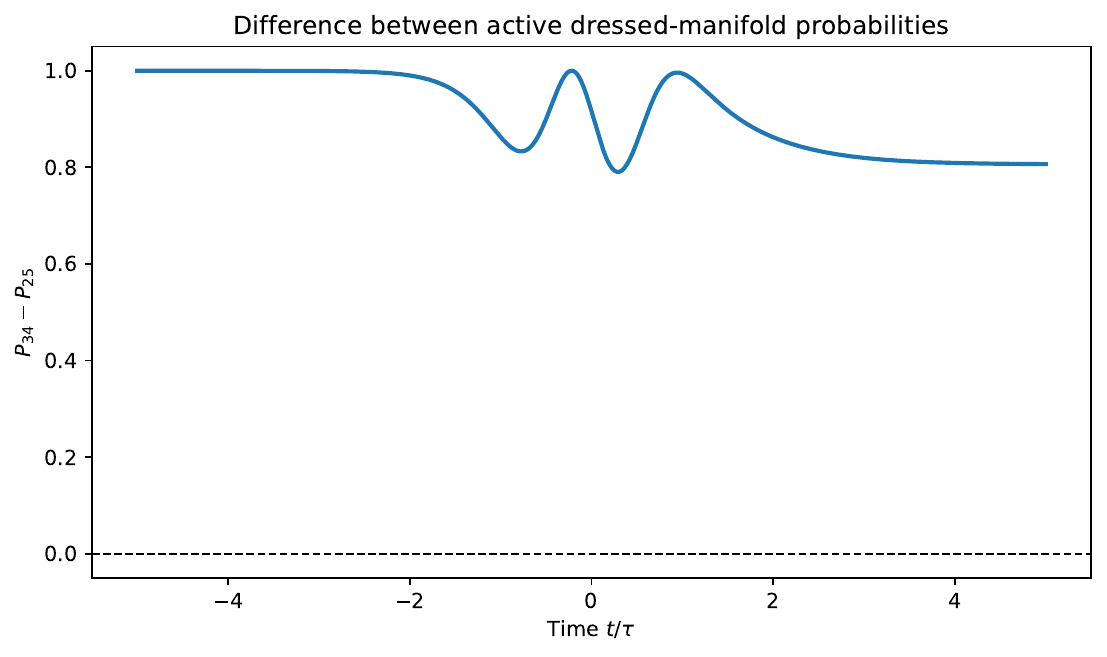}
\caption{
Difference between the populations of the central and neighboring dressed
manifolds.
The temporary reduction of
$\Delta P$
indicates coherent exchange between the two manifolds during the pulse
overlap.
}
\label{fig:darkbrightdifference}
\end{figure}

To emphasize this structure,
Fig.~\ref{fig:darkbrightdifference}
presents the difference between the populations of the central and
neighboring dressed manifolds,
\begin{equation}
\Delta P
=
P_{34}
-
P_{25},
\end{equation}
where 
$P_{34}=P_3+P_4,
%\qquad
P_{25}=P_2+P_5.$ 
The reduction of
$\Delta P$
during the pulse-overlap interval demonstrates that population is exchanged
coherently between these two manifolds while the outer manifold remains almost
inactive.
These observations suggest that the oscillatory transfer is governed neither
by interference between bare-state pathways nor by evolution of isolated
dressed eigenstates.
Instead, the numerical evidence points toward coherent dynamics of coupled
dressed manifolds.
To formulate this mechanism quantitatively, we next introduce a
gauge-invariant projector formalism that allows the collective manifold
dynamics to be analyzed independently of the particular basis chosen within
each degenerate subspace.
%%%%%%%%%%%%%%%%%%%%%%%%%%%%%%%%%%%%%%%%%%%%%%%%%%%%%%%%%%%%%%%%%%%%%%%%%%%%%%
\subsection{Gauge-Invariant Projector Description of Collective Dressed Manifolds}

The results of the previous subsection demonstrate that the instantaneous
dressed states do not participate independently in the evolution. Instead, the
population remains concentrated within three well-defined spectral clusters,
consisting of the pairs
$(\Phi_1,\Phi_6)$,
$(\Phi_2,\Phi_5)$,
and
$(\Phi_3,\Phi_4)$.
%The numerical evidence therefore suggests that the physically relevant dynamical objects are the corresponding dressed manifolds rather than the individual eigenstates.

A direct analysis of dressed states is, however, not entirely
satisfactory~\cite{Kato1950,Berry1984,WilczekZee1984,Avron1987}. Within a degenerate manifold, the instantaneous
eigenvectors are not uniquely defined. Small numerical perturbations may
rotate the basis inside the manifold without affecting the physical
wavefunction. Consequently, quantities associated with individual eigenvectors
may depend on the particular gauge chosen during numerical diagonalization.

To eliminate this ambiguity, we introduce projection operators onto the three
instantaneous dressed manifolds,
\begin{equation}
\hat P_D
=
\sum_{i=3,4}
|\Phi_i\rangle
\langle\Phi_i|,
\label{eq:PD}
\end{equation}
\begin{equation}
\hat P_B
=
\sum_{i=2,5}
|\Phi_i\rangle
\langle\Phi_i|,
\label{eq:PB}
\end{equation}
\begin{equation}
\hat P_O
=
\sum_{i=1,6}
|\Phi_i\rangle
\langle\Phi_i|,
\label{eq:PO}
\end{equation}
where the subscripts denote the dominant dark,
neighboring bright,
and outer manifolds,
respectively.

Because the dressed states constitute a complete orthonormal basis,
\begin{equation}
\sum_{i=1}^{6}
|\Phi_i\rangle\langle\Phi_i|
=
\mathbb I,
\end{equation}
the three projectors satisfy
\begin{equation}
\hat P_D
+
\hat P_B
+
\hat P_O
=
\mathbb I.
\label{eq:resolution}
\end{equation}
Furthermore,
\begin{equation}
\hat P_\alpha^2
=
\hat P_\alpha,
\qquad
\hat P_\alpha\hat P_\beta
=
0,
\qquad
(\alpha\neq\beta),
\end{equation}
demonstrating that the three manifolds form mutually orthogonal invariant
subspaces of the instantaneous Hilbert space.

%%%%%%%%%%%%%%%%%%%%%%%%%%%%%%%%%%%%%%%%%%%%%%%%%%%%%%%%%%%%%%%%%%%%%%%%%%%

The propagated wavefunction may therefore be decomposed uniquely into the
three manifold components,
\begin{equation}
|\Psi(t)\rangle
=
|\Psi_D(t)\rangle
+
|\Psi_B(t)\rangle
+
|\Psi_O(t)\rangle,
\label{eq:decomposition}
\end{equation}
where
\begin{equation}
|\Psi_D\rangle
=
\hat P_D|\Psi\rangle,
\label{eqn: Dark}
\end{equation}
\begin{equation}
|\Psi_B\rangle
=
\hat P_B|\Psi\rangle,
\label{eqn: Bright}
\end{equation}
\begin{equation}
|\Psi_O\rangle
=
\hat P_O|\Psi\rangle.
\label{eqn: Outer}
\end{equation}
Unlike the coefficients of individual dressed eigenstates, these projected
wavefunctions are invariant under any unitary rotation performed inside a
given manifold. They therefore represent genuine physical observables rather
than quantities depending on an arbitrary choice of basis.

The corresponding manifold populations are
\begin{equation}
P_D(t)
=
\langle\Psi_D|\Psi_D\rangle,
\end{equation}
\begin{equation}
P_B(t)
=
\langle\Psi_B|\Psi_B\rangle,
\end{equation}
\begin{equation}
P_O(t)
=
\langle\Psi_O|\Psi_O\rangle,
\end{equation}
which satisfy
\begin{equation}
P_D
+
P_B
+
P_O
=
1.
\label{eq:population_sum}
\end{equation}

%%%%%%%%%%%%%%%%%%%%%%%%%%%%%%%%%%%%%%%%%%%%%%%%%%%%%%%%%%%%%%%%%%%%%%%%%%%

\begin{figure}[t]

\centering

\includegraphics[width=\columnwidth]{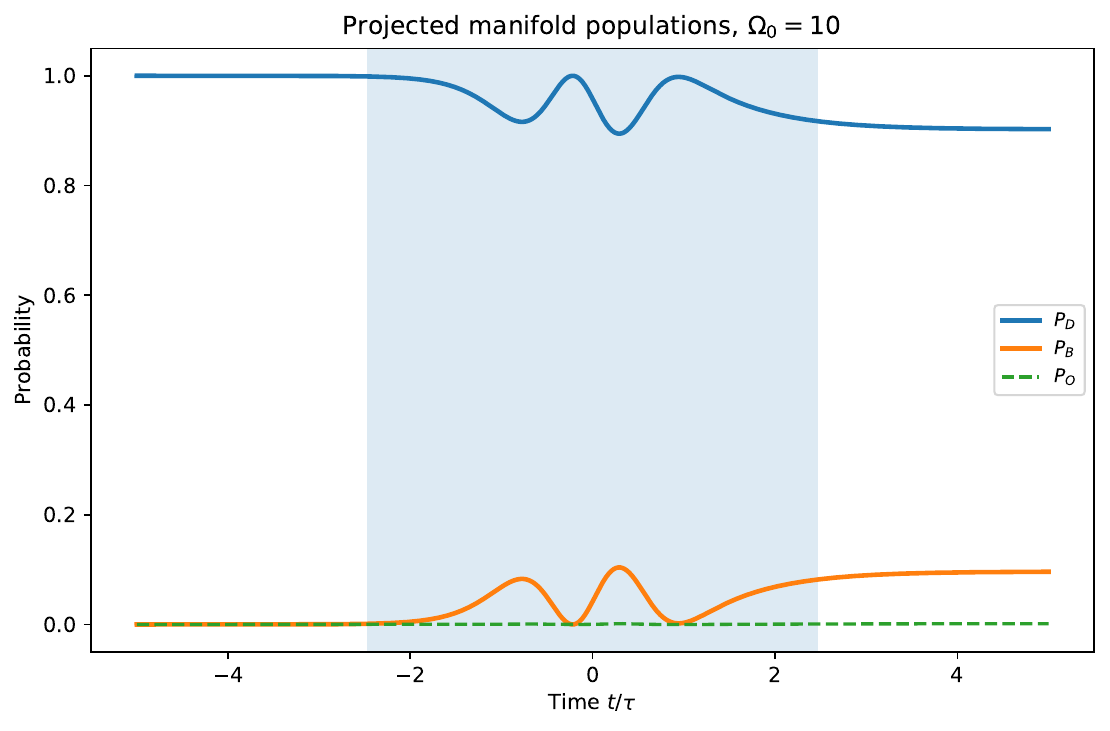}

\caption{
Gauge-invariant populations of the three dressed manifolds obtained from the
projector formalism.
The wavefunction remains predominantly inside the dark manifold while the
neighboring bright manifold acquires a temporary population during the pulse
overlap. The outer manifold remains almost completely unoccupied throughout
the evolution. The blue shaded background marks the pulse-overlap region.
}

\label{fig:ProjectedPopulations}

\end{figure}

Figure~\ref{fig:ProjectedPopulations}
shows the evolution of these three populations obtained from the numerical
propagation  with the blue shaded region indicating the temporal interval of strongest overlap between the pump and Stokes pulses.
Three distinct stages of the dynamics can be identified. Initially,
\[
P_D\simeq1,
\]
indicating that the collective wavefunction is almost entirely confined to the
dark manifold. As the pump and Stokes pulses begin to overlap,
population is partially  transferred coherently into the neighboring bright manifold.
The corresponding decrease of
$P_D$
is accompanied by an increase of
$P_B$,
whereas
\[
P_O\approx0
\]
during the entire evolution. Finally,
at the end of pulse overlap,
the transferred population returns almost completely to the dark manifold,
which again dominates the wavefunction at the end of the STIRAP sequence.

These results establish an important physical conclusion.
The neighboring bright manifold is not merely a weak perturbative correction
to an otherwise isolated dark-state evolution.
Instead, it actively participates in the collective dynamics by temporarily
accepting population from the dark manifold before returning it later in the
pulse sequence.

\begin{figure}[t]
\centering
\includegraphics[width=\columnwidth]{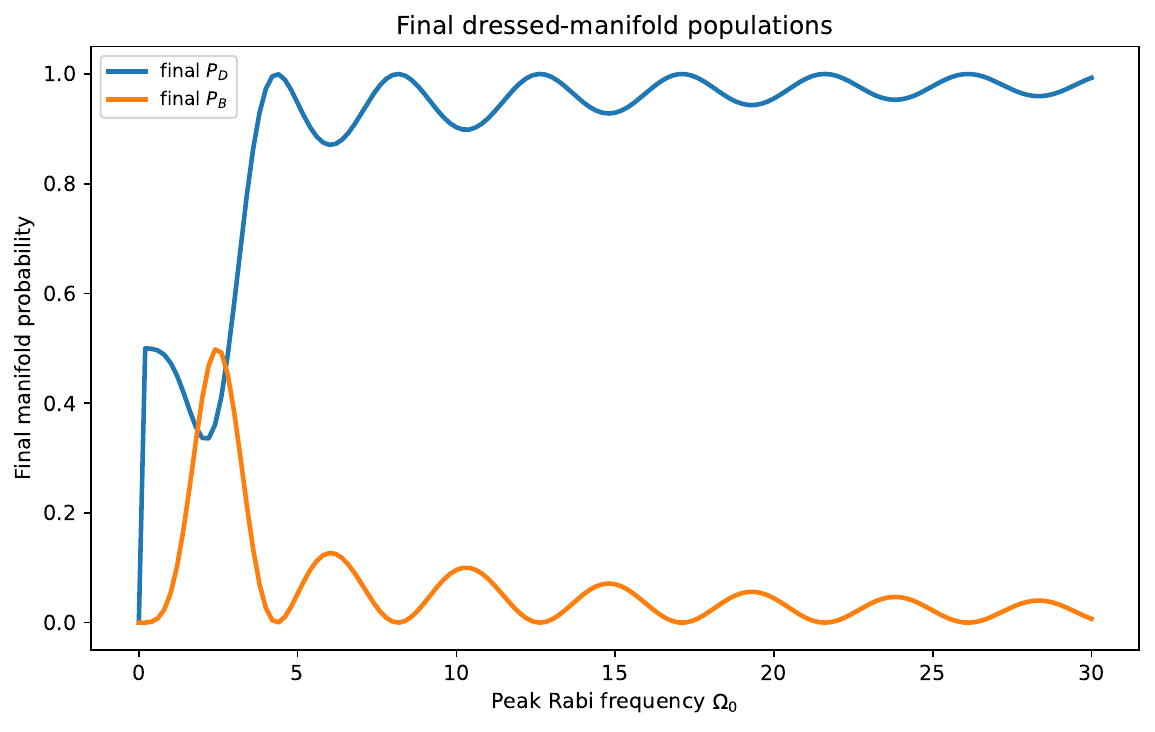}
\caption{Final populations of the projected dark and neighboring bright
manifolds as functions of the peak Rabi frequency $\Omega_0$.  Above the
transfer threshold, the dark-manifold population remains dominant but exhibits
an oscillatory modulation.  The complementary maxima of $P_B^{(f)}$ identify
those pulse amplitudes for which the wavefunction does not return completely
to the dark manifold by the end of the pulse sequence.}
\label{fig:FinalManifoldsVsOmega}
\end{figure}

The dependence of the final manifold populations on the peak Rabi frequency
provides a complementary view of the oscillatory transfer.  Figure~\ref{fig:FinalManifoldsVsOmega}
shows the populations remaining in the projected dark and bright manifolds at
the end of the pulse sequence:
%\begin{equation}
$P_D^{(f)}(\Omega_0), $ %=P_D(t_f;\Omega_0),
%\qquad
and $P_B^{(f)}(\Omega_0).$ %=P_B(t_f;\Omega_0).
%\label{eq:final_manifold_populations}
%\end{equation}

At small pulse amplitudes the adiabatic passage is not established and a
substantial fraction of the wavefunction can remain in the neighboring bright
manifold.  Once the collective-transfer threshold is crossed,
$P_D^{(f)}$ approaches unity, but it does so nonmonotonically.  Its minima are
accompanied by maxima of $P_B^{(f)}$, demonstrating that the oscillatory
response is associated with incomplete dark--bright--dark population return.  As
$\Omega_0$ increases, the bright-manifold residue decreases and the modulation
becomes progressively weaker, consistent with improved adiabatic isolation of
the dark manifold.  This scan therefore connects the time-resolved population
exchange in Fig.~\ref{fig:ProjectedPopulations} directly to the oscillatory
pulse-area dependence observed in Sec.~III.

The projector formalism therefore reveals that the collective oscillatory
transfer is fundamentally a problem of communication between dressed
manifolds.
To quantify this communication, we next introduce a generalized
wavefunction-projected nonadiabatic coupling that plays the role analogous to
the familiar STIRAP mixing rate in the conventional three-level problem.
%%%%%%%%%%%%%%%%%%%%%%%%%%%%%%%%%%%%%%%%%%%%%%%%%%%%%%%%%%%%%%%%%%%%%%%%%%%%%%
\subsection{Wavefunction-Projected Nonadiabatic Coupling Between Dressed Manifolds}

The projector analysis presented above establishes that the collective
wavefunction evolves predominantly inside the dark manifold while undergoing
temporary population exchange with the neighboring bright manifold. The
remaining question is therefore to identify the physical mechanism
responsible for this inter-manifold communication.

For a conventional three-level STIRAP system, the dynamics is governed by a
single instantaneous dark state~\cite{Bergmann1998,Vitanov2017,Vasilev2009},
\begin{equation}
|D(t)\rangle
=
\cos\theta(t)|1\rangle
-
\sin\theta(t)|3\rangle,
\end{equation}
whose evolution is controlled by the mixing angle
\begin{equation}
\tan\theta(t)
=
\frac{\Omega_P(t)}
{\Omega_S(t)}.
\end{equation}
The corresponding nonadiabatic coupling is proportional to~\cite{Bergmann1998,Vitanov2017}
%\begin{equation}
$\dot\theta(t),$ 
%\end{equation}
which determines the probability of leaving the dark state.

The present problem is more complex.
Instead of a single dark eigenstate,
the collective system evolves inside a multidimensional dressed manifold, for which projector and non-Abelian formulations provide a natural gauge-independent language~\cite{Kato1950,WilczekZee1984,Avron1987}.
Consequently,
no unique mixing angle exists,
and the conventional STIRAP description cannot be generalized directly.

%%%%%%%%%%%%%%%%%%%%%%%%%%%%%%%%%%%%%%%%%%%%%%%%%%%%%%%%%%%%%%%%%%%%%%%%%%%

To characterize the communication between manifolds,
we work with the projected wavefunctions defined in Eqs(\ref{eqn: Dark}-\ref{eqn: Outer}) 
%\[
%|\Psi_D(t)\rangle
%=
%\hat P_D|\Psi(t)\rangle,
%\] \[ |\Psi_B(t)\rangle
%=
%\hat P_B|\Psi(t)\rangle  \]
%and
%\[
%|\Psi_O(t)\rangle
%=
%\hat P_O|\Psi(t)\rangle,
%\]
and define the scalar coupling
\begin{equation}
\Gamma_{DB}(t)
=
\left|
\left\langle
\Psi_B(t)
\middle|
\frac{d}{dt}
\Psi_D(t)
\right\rangle
\right|.
\label{eq:GammaDB}
\end{equation}
Similarly,
\begin{equation}
\Gamma_{DO}(t)
=
\left|
\left\langle
\Psi_O(t)
\middle|
\frac{d}{dt}
\Psi_D(t)
\right\rangle
\right|.
\label{eq:GammaDO}
\end{equation}
%Unlike matrix-valued non-Abelian connections,
Eqs.~(\ref{eq:GammaDB}) and
(\ref{eq:GammaDO})
measure directly the communication between the occupied manifold components of
the propagated wavefunction.
Because they are constructed from projector-defined wavefunctions,
they are invariant under unitary rotations within each manifold and therefore
constitute physically meaningful observables.

%%%%%%%%%%%%%%%%%%%%%%%%%%%%%%%%%%%%%%%%%%%%%%%%%%%%%%%%%%%%%%%%%%%%%%%%%%%

\begin{figure}[t]

\centering

\includegraphics[width=\columnwidth]{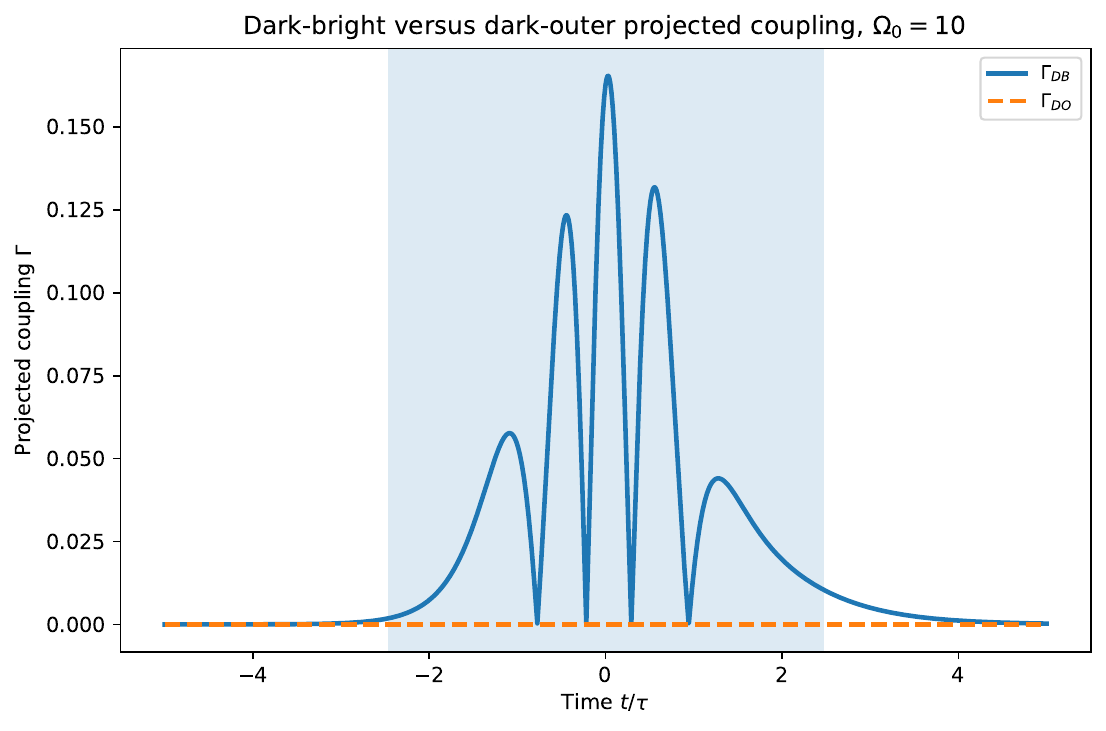}

\caption{
Wavefunction-projected nonadiabatic coupling
$\Gamma_{DB}(t)$
between the dark and neighboring bright manifolds.
The coupling is strongly localized in the pulse-overlap region. Projected coupling
$\Gamma_{DO}$ demonstrates that
communication of the dark manifold  with the neighboring bright manifold dominates throughout the
entire evolution.
}

\label{fig:GammaDB}

\end{figure}

Figure~\ref{fig:GammaDB}
shows the evolution of
$\Gamma_{DB}(t)$.
The coupling exhibits a pronounced maximum during the interval where the pump
and Stokes pulses overlap most strongly.
Outside this region,
$\Gamma_{DB}$
rapidly approaches zero,
indicating that the dark manifold becomes effectively decoupled from the
remaining spectrum both before and after the interaction.
%%%%%%%%%%%%%%%%%%%%%%%%%%%%%%%%%%%%%%%%%%%%%%%%%%%%%%%%%%%%%%%%%%%%%%%%%%%
To determine whether this communication is unique to the neighboring bright
manifold,
%Fig.~\ref{fig:GammaComparison}
%in comparison, 
%$\Gamma_{DB}(t)$
%with
$\Gamma_{DO}(t)$ is also shown in this Figure. The coupling to the outer manifold remains more than an order of magnitude
smaller throughout the propagation.
The collective dynamics therefore exhibits a remarkable degree of selectivity:
the dark manifold communicates almost exclusively with the neighboring bright
manifold while remaining effectively isolated from the rest of the spectrum.

%%%%%%%%%%%%%%%%%%%%%%%%%%%%%%%%%%%%%%%%%%%%%%%%%%%%%%%%%%%%%%%%%%%%%%%%%%%

The significance of this communication becomes evident by comparing the
projected coupling with the instantaneous dressed-manifold energy separation.

\begin{figure}[t]

\centering

\includegraphics[width=\columnwidth]{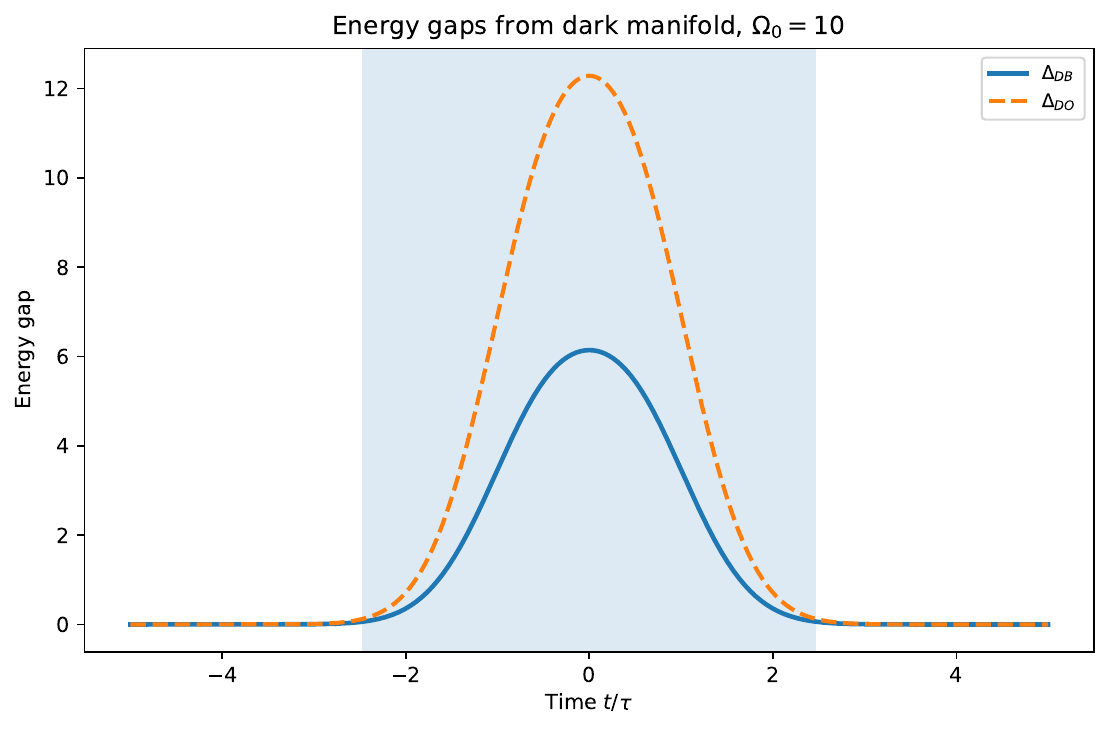}

\caption{
Instantaneous energy separation between the dark and neighboring
bright manifolds.
The energy gap remains finite throughout the pulse sequence.
}

\label{fig:Gap}

\end{figure}

Figure~\ref{fig:Gap}
shows the energy gap between the dark and neighboring bright
manifolds.
The gap never closes during the evolution.
%Instead, its minimum occurs near the same temporal region in which $\Gamma_{DB}(t)$ reaches its maximum.
This observation has an important physical consequence. The population transfer between manifolds is {\em not} caused by a collapse of
the adiabatic spectrum.
Rather,
it originates from localized nonadiabatic communication between neighboring
spectral manifolds while the overall adiabatic ordering of the dressed states
remains intact.

A useful dimensionless measure of this competition is the projected
nonadiabaticity parameter
\begin{equation}
\eta_{DB}(t;\Omega_0)
=
\frac{\Gamma_{DB}(t;\Omega_0)}{\Delta_{DB}(t;\Omega_0)},
\label{eq:etaDB}
\end{equation}
where $\Delta_{DB}$ is the minimum instantaneous energy separation between the
dark and neighboring bright manifolds.  This quantity is the reciprocal of a
Massey-type adiabaticity parameter: $\eta_{DB}\ll1$ corresponds to adiabatic
isolation, whereas values of order unity or larger signal appreciable
nonadiabatic communication~\cite{Shore2011,Vitanov2017}.  To characterize an
entire pulse sequence, we evaluate both the maximum value and the temporal mean
within the active pulse-overlap window,
\begin{eqnarray}
&\eta_{DB}^{\max}(\Omega_0)
=
\max_{t\in{t_d}}\eta_{DB}(t;\Omega_0),\nonumber\\
&\overline{\eta}_{DB}(\Omega_0)
=
\frac{1}{t_d}
\int_{t_d}\eta_{DB}(t;\Omega_0)\,dt.
\label{eq:etaDBmeasures}
\end{eqnarray}

\begin{figure}[t]
\centering
\includegraphics[width=\columnwidth]{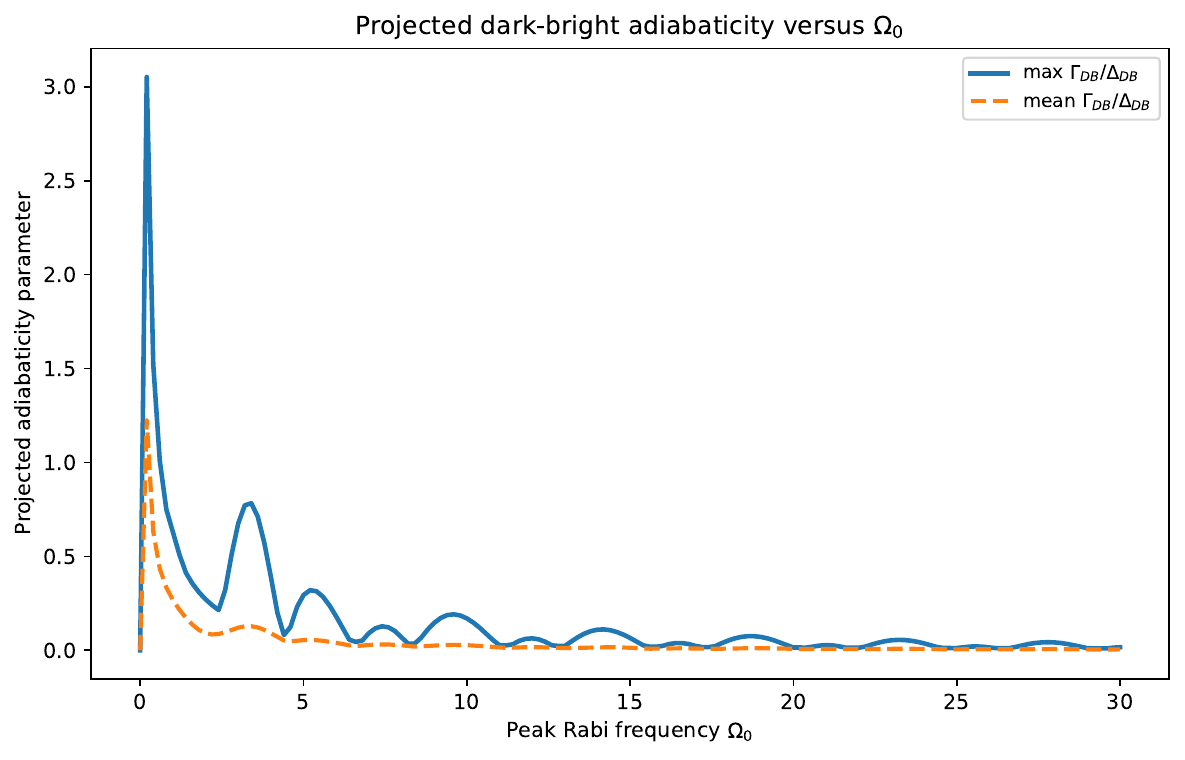}
\caption{Maximum and active-window mean of the projected dark--bright
nonadiabaticity parameter $\eta_{DB}=\Gamma_{DB}/\Delta_{DB}$ as functions of
the peak Rabi frequency.  The large values at weak driving mark the breakdown
of adiabatic isolation.  With increasing $\Omega_0$, both measures decrease,
although the maximum retains oscillatory peaks associated with localized
bursts of dark--bright manifold communication.}
\label{fig:ProjectedAdiabaticityVsOmega}
\end{figure}
Figure~\ref{fig:ProjectedAdiabaticityVsOmega} shows that the weak-field region
is strongly nonadiabatic: the maximum ratio can exceed unity because the
available energy separation is insufficient to suppress the time-dependent
manifold coupling.  Beyond the collective-transfer threshold, the mean value
falls rapidly and remains small, confirming that the evolution is globally
adiabatic over most of the pulse sequence.  The maximum value, however,
exhibits a sequence of diminishing peaks.  These peaks show that globally
adiabatic evolution can still contain short, coherent episodes of enhanced
communication between the neighboring manifolds.  Their decreasing envelope
with increasing $\Omega_0$ is consistent with the simultaneous suppression of
the residual final bright-manifold population shown in
Fig.~\ref{fig:FinalManifoldsVsOmega}.  Thus the two curves provide 
consistent evidence: stronger driving improves overall adiabaticity, while the
remaining oscillatory structure is generated by localized dark--bright
coupling events in the pulse-overlap interval.
%%%%%%%%%%%%%%%%%%%%%%%%%%%%%%%%%%%%%%%%%%%%%%%%%%%%%%%%%%%%%%%%%%%%%%%%%%%

\begin{figure}[t]

\centering

\includegraphics[width=\columnwidth]{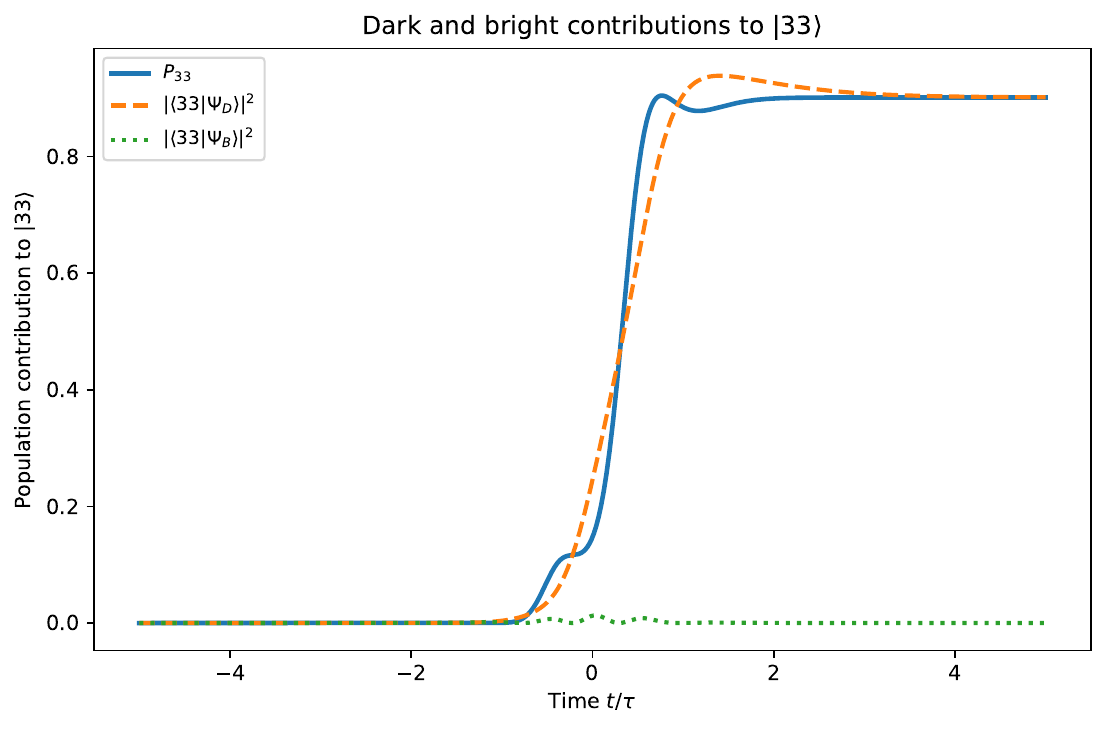}

\caption{
Decomposition of the target-state population into contributions from the
projected dark and neighboring bright manifolds.
}

\label{fig:DarkContribution}

\end{figure}

The final step is to determine how this transient communication contributes to
the transfer into the doubly excited Rydberg state. Figure~\ref{fig:DarkContribution}
decomposes the final population
%\[
$P_{33}(t)$ 
%\]
into contributions originating from the projected dark and bright manifold
components
%\[ 
$|\langle33|\Psi_D\rangle|^2$ 
%\]
and
%\[
 $|\langle33|\Psi_B\rangle|^2.$ 
%\]

Although the neighboring bright manifold acquires a measurable transient
population during the pulse-overlap interval,
its direct contribution to the
final population of
$|33\rangle$
remains small.
Instead,
the transferred population is recovered almost entirely within the projected
dark manifold before completion of the STIRAP sequence.

The collective oscillatory transfer is therefore governed by the following
mechanism: 
The wavefunction evolves predominantly inside a two-dimensional dark manifold,
temporarily communicates with the neighboring bright manifold through a
localized nonadiabatic coupling,
and subsequently returns to the dark manifold,
which carries the final population transferred to the doubly excited Rydberg
state.

This dressed-manifold mechanism provides a physically transparent explanation
of the oscillatory collective STIRAP dynamics observed in Sec.~III and forms
the foundation for understanding the influence of Rydberg--Rydberg
interactions discussed in the following section.
%%%%%%%%%%%%%%%%%%%%%%%%%%%%%%%%%%%%%%%%%%%%%%%%%%%%%%%%%%%%%%%%%%%%%%%%%%%%%%
\section{Compensation of Rydberg--Rydberg Interaction by Chirped STIRAP}

\subsection{Interaction-Induced Detuning of the Collective Raman Resonance}

The dressed-manifold mechanism developed in the previous section explains the
origin of the oscillatory collective transfer in the absence of
Rydberg--Rydberg interactions.
We now investigate how this mechanism is modified when the doubly excited
Rydberg state is shifted by the interaction energy
$V_{\rm RR}$~\cite{Jaksch2000,Lukin2001,Saffman2010,Petrosyan2017,Saffman2020}.

Within the effective Hamiltonian introduced in Sec.~II, the interaction
contributes only to the collective pair state 
%\begin{equation}
$|33\rangle,$ 
%\end{equation}
whose equation of motion may be written as
\begin{equation}
\frac{d a_{33}}{dt}
=
-i
\frac{\Omega_S(t)}{\sqrt2}
a_{23}
-i
\left[
2\delta(t)
+
V_{\rm RR}
\right]
a_{33},
\label{eq:a33}
\end{equation}
where the interaction appears as an additional detuning of the final Raman
transition.

Equation (\ref{eq:a33}) immediately shows that the effective resonance
condition is no longer determined solely by the two-photon detuning.
Instead, resonance occurs whenever
\begin{equation}
2\delta(t_{\rm res})
+
V_{\rm RR}
=
0.
\label{eq:pairres}
\end{equation}
Thus, a repulsive interaction
($V_{\rm RR}>0$)
shifts the collective Raman resonance away from its noninteracting position.
If the chirp is absent,
\begin{equation}
\delta(t)=\delta_0,
\end{equation}
the resonance condition becomes
\begin{equation}
2\delta_0
+
V_{\rm RR}
=
0,
\end{equation}
showing that a nonzero interaction inevitably introduces a mismatch whenever
the initial two-photon detuning is chosen to be  
%\[
$\delta_0=0.$ 
%\]

Consequently, increasing the interaction suppresses the population transfer
unless sufficiently large Rabi frequencies restore the dynamics through
power broadening, as in the standard blockade picture~\cite{Urban2009,Gaetan2009,Wilk2010,Isenhower2010}.

The situation changes qualitatively when linear chirp is introduced.
Using
\begin{equation}
\delta(t)
=
\delta_0
-
2\alpha
(t-t_P),
\label{eq:chirpdelta}
\end{equation}
the resonance condition becomes
\begin{equation}
2\delta_0
-
4\alpha
(t_{\rm res}-t_P)
+
V_{\rm RR}
=
0,
\end{equation}
yielding the general expression
\begin{equation}
%\boxed{
\alpha
=
\frac{
2\delta_0
+
V_{\rm RR}
}
{
4
(t_{\rm res}-t_P)
}.
%}
\label{eq:alphageneral}
\end{equation}
Equation (\ref{eq:alphageneral}) constitutes a general analytical design rule
for chirped collective STIRAP.
Rather than compensating the interaction through increased laser intensity,
the resonance may be restored dynamically by choosing a chirp that sweeps the
collective pair state through resonance at the desired instant.

A particularly important situation corresponds to zero initial two-photon
detuning, 
%\begin{equation}
$\delta_0=0,$ 
%\end{equation}
for which Eq.~(\ref{eq:alphageneral}) reduces to
\begin{equation}
%\boxed{
\alpha
=
\frac{
V_{\rm RR}
}
{
4
(t_{\rm res}-t_P)
}.
%}
\label{eq:alphazero}
\end{equation}
In the present work the pump pulse reaches its maximum at
$t=t_P$,
whereas the Stokes pulse precedes it by
%\[
$t_d=0.75\tau.$ 
%\]
The strongest collective coupling therefore occurs near the center of the
pulse-overlap interval,
\begin{equation}
t_{\rm ov}
=
t_P
-
\frac{t_d}{2}.
\end{equation}
Choosing the resonance to coincide with this interval,
\[
t_{\rm res}
=
t_{\rm ov},
\]
gives
\begin{equation}
%\boxed{
\alpha
=
-
\frac{
V_{\rm RR}
}
{
2t_d
}.
%}
\label{eq:optimalchirp}
\end{equation}
For the pulse sequence employed throughout this work, 
%\[
$t_d
=
0.75\tau,$ 
%\]
so that
\begin{equation}
%\boxed{
\alpha
=
-
\frac{
V_{\rm RR}
}
{
1.5\tau
}.
%}
\label{eq:practical}
\end{equation}
Equation (\ref{eq:practical}) predicts the chirp that restores the collective
Raman resonance near the region where the pump and Stokes pulses overlap most
strongly. As shown below, numerical simulations demonstrate that this simple
analytical estimate successfully predicts the onset of efficient collective
population transfer in the interacting system.

%%%%%%%%%%%%%%%%%%%%%%%%%%%%%%%%%%%%%%%%%%%%%%%%%%%%%%%%%%%%%%%%%%%%%%%%%%%%%%
\subsection{Numerical Verification of Chirp Compensation}

The analytical resonance condition derived in the previous subsection predicts
that a suitable linear chirp can compensate the interaction-induced shift of
the doubly excited Rydberg state. We now verify this prediction by solving the
complete time-dependent Schrödinger equation for the interacting six-state
system.

Throughout this section the Rydberg--Rydberg interaction strength is fixed at
%\[
$V_{\rm RR}=5 [\omega],$ 
%\]
while all remaining pulse parameters are identical to those employed in the
noninteracting calculations of Sec.~III.

%%%%%%%%%%%%%%%%%%%%%%%%%%%%%%%%%%%%%%%%%%%%%%%%%%%%%%%%%%%%%%%%%%%%%%%%%%%

\begin{figure}[t]

\centering

\includegraphics[width=\columnwidth]{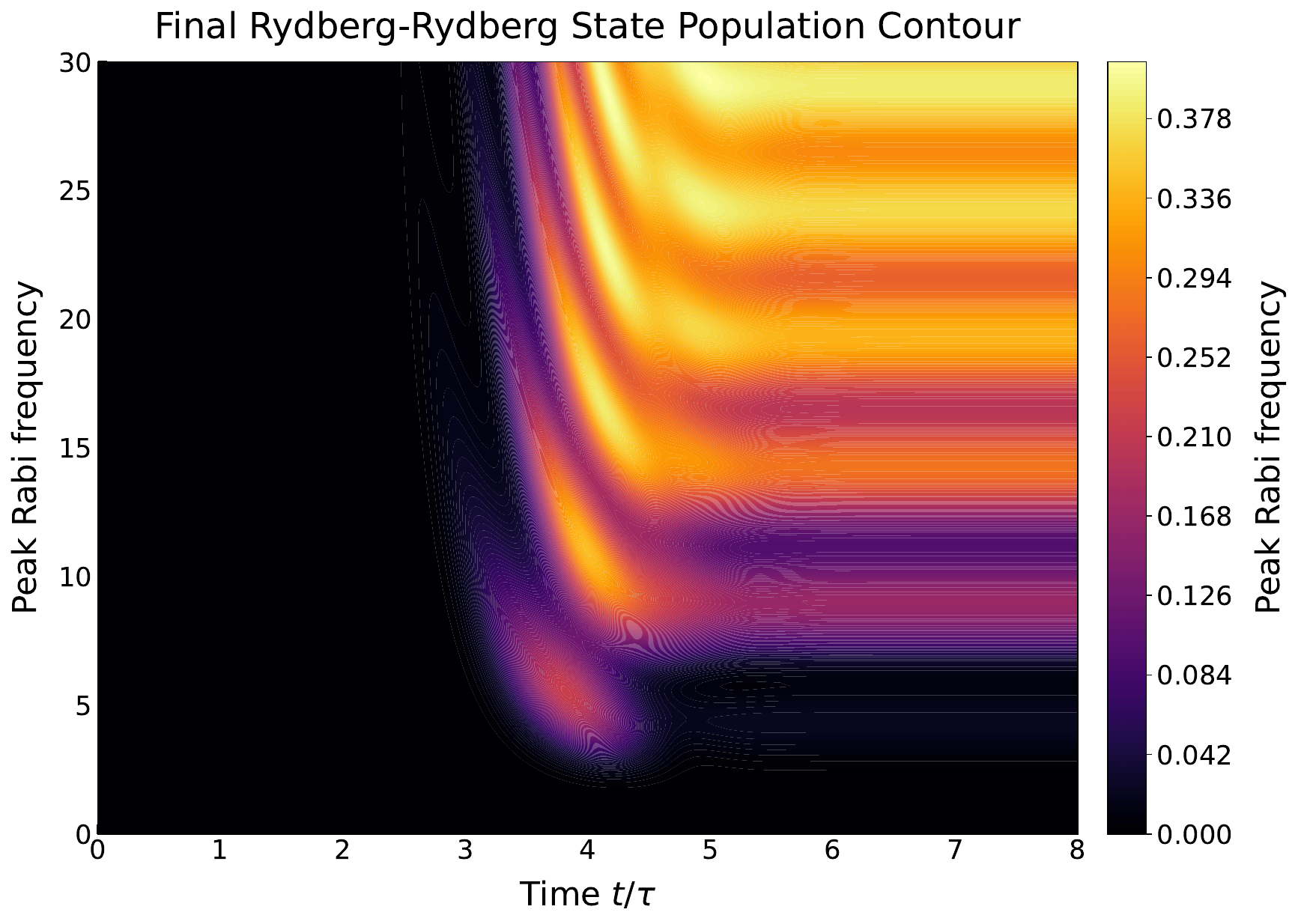}

\caption{
Contour plot of the population transferred to the doubly excited Rydberg
state for
$V_{\rm RR}=5$
without chirp.
The interaction shifts the collective Raman resonance, suppressing efficient
population transfer for small and moderate pulse amplitudes.
}

\label{fig:V5NoChirp}

\end{figure}

Figure~\ref{fig:V5NoChirp}
shows the contour plot of the population transferred to the doubly excited
Rydberg state,
$P_{33}(t),$ 
for zero chirp,   
$\alpha=0.$ 

The interaction shifts the pair state away from the two-photon resonance,
thereby reducing the transfer efficiency.
Consequently,  excitation up to 0.4 occurs only for sufficiently large pulse
areas where strong laser coupling partially compensates the interaction
through increased spectral broadening.

Nevertheless, the oscillatory dependence on the pulse area remains clearly
visible.
This observation immediately indicates that the oscillatory transfer mechanism
identified in Sec.~IV survives the presence of the interaction.

%%%%%%%%%%%%%%%%%%%%%%%%%%%%%%%%%%%%%%%%%%%%%%%%%%%%%%%%%%%%%%%%%%%%%%%%%%%

\begin{figure}[t]

\centering

\includegraphics[width=\columnwidth]%{V5_Chirp.pdf}
{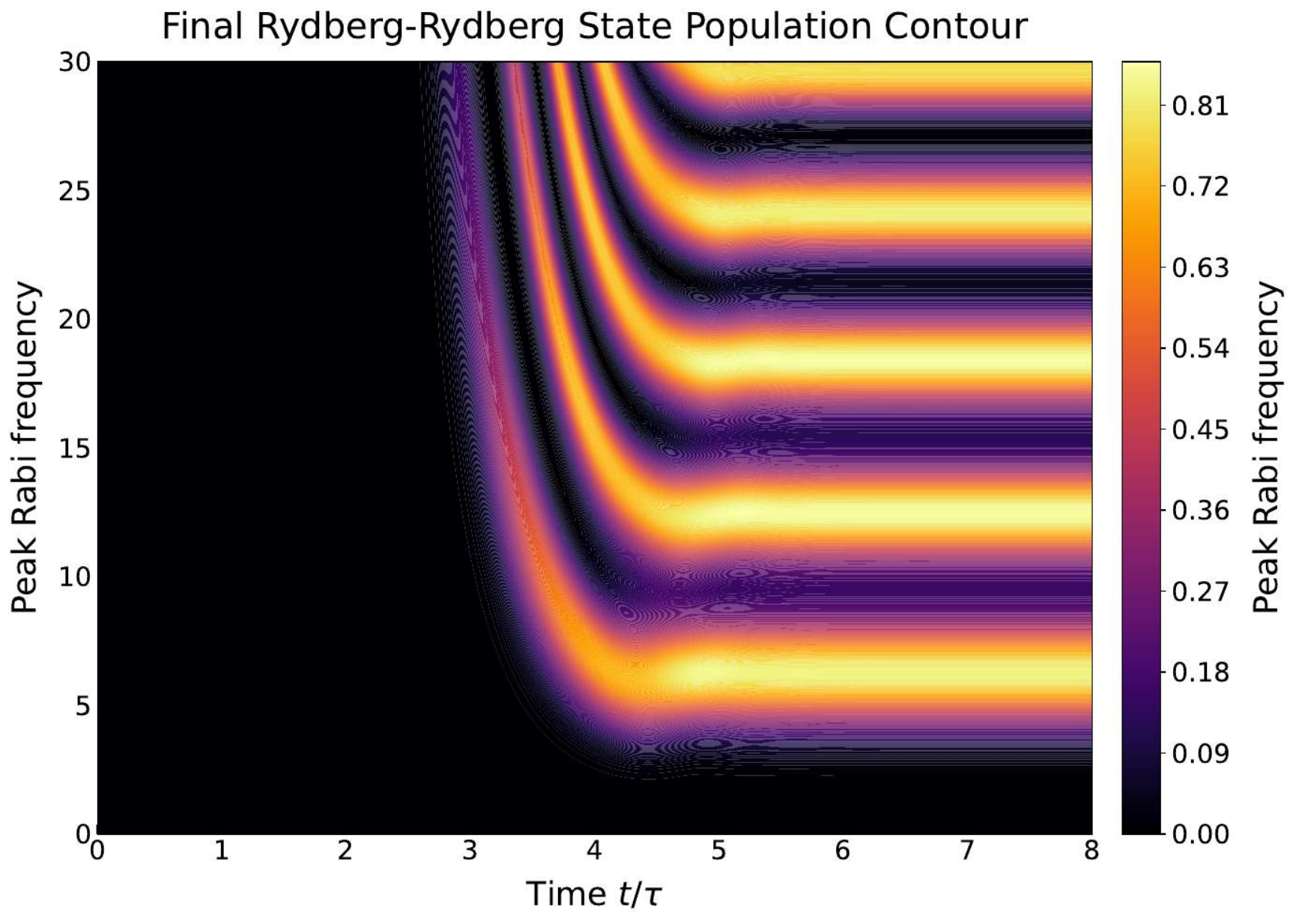}
\caption{
Contour plot of the doubly excited Rydberg-state population obtained using
the analytically predicted chirp
$\alpha=-V_{\rm RR}/1.5$.
The oscillatory transfer bands are preserved while efficient population
transfer is restored at considerably lower pulse amplitudes.
}

\label{fig:V5Chirp}

\end{figure}

Figure~\ref{fig:V5Chirp}
shows the corresponding calculation performed using the analytical chirp in Eq.(\ref{eq:practical}), 
which for
\[
V_{\rm RR}=5,
\qquad
\tau=1,
\]
gives
\[
\alpha=-3.33.
\]
The effect of chirping is immediately apparent. First, the onset of efficient transfer shifts toward substantially smaller
Rabi frequencies.
Second, the alternating oscillatory bands remain clearly visible.
Finally, the maximum transferred population approaches unity over a broad
range of pulse amplitudes despite the presence of the interaction.

%%%%%%%%%%%%%%%%%%%%%%%%%%%%%%%%%%%%%%%%%%%%%%%%%%%%%%%%%%%%%%%%%%%%%%%%%%%

The comparison between
Figs.~\ref{fig:V5NoChirp}
and
\ref{fig:V5Chirp}
leads to an important physical conclusion: The chirp does not eliminate the oscillatory collective dynamics.
Instead, it restores the instantaneous Raman resonance while preserving the
same dressed-manifold transfer mechanism identified for the noninteracting system.

In other words, the interaction modifies the resonance condition but does not
change the underlying topology of the collective dressed-state evolution.

%%%%%%%%%%%%%%%%%%%%%%%%%%%%%%%%%%%%%%%%%%%%%%%%%%%%%%%%%%%%%%%%%%%%%%%%%%%

The present results therefore suggest that the Rydberg--Rydberg interaction determines the location of the collective
Raman resonance, whereas the coherent oscillatory transfer is governed by the
communication between neighboring dressed manifolds.
Frequency chirping provides an efficient mechanism for compensating the former
without altering the latter.

This separation of roles explains why the oscillatory bands persist in the
interacting system while their position in parameter space is shifted by the
interaction and subsequently restored by the appropriately chosen chirp.

The numerical agreement with the analytical resonance condition derived in
Eq.~(\ref{eq:practical}) demonstrates that the simple chirp-compensation rule
captures the essential physics of the interacting collective STIRAP process
despite the considerably more complex six-state dynamics.

%%%%%%%%%%%%%%%%%%%%%%%%%%%%%%%%%%%%%%%%%%%%%%%%%%%%%%%%%%%%%%%%%%%%%%%%%%%%%%
\subsection{Physical Picture of Interaction Compensation}

The analytical and numerical results presented above lead to a unified physical
picture of collective chirped STIRAP in the presence of
Rydberg--Rydberg interactions.

In the absence of interactions, the collective transfer cannot be understood
within the conventional three-level STIRAP paradigm.
Although the six-state Hamiltonian contains multiple bare-state excitation
paths connecting the initial and target states, the bare-state populations
demonstrate that the oscillatory dependence of the transfer efficiency is not
simply the consequence of interference between alternative excitation
pathways.

Instead, the dynamics is governed by the instantaneous dressed-state
structure.
The wavefunction evolves predominantly inside a two-dimensional dark manifold,
while temporarily exchanging population with a neighboring bright manifold
during the pulse-overlap interval.
The corresponding communication is quantified by the localized
wavefunction-projected nonadiabatic coupling
$\Gamma_{DB}(t)$.
After this transient exchange, the wavefunction returns almost completely to
the dark manifold, which carries essentially the entire final population
transferred to the doubly excited Rydberg state.

The introduction of the Rydberg--Rydberg interaction does not modify this
basic mechanism.
Instead, it shifts the energy of the collective pair state
$|33\rangle$,
thereby changing the instantaneous Raman resonance condition.
Consequently, the onset of efficient transfer moves toward larger pulse areas,
while the oscillatory structure itself remains remarkably robust.

The role of the chirp is fundamentally different from that of the interaction.
Rather than altering the dressed-manifold dynamics, the chirp compensates the
interaction-induced energy shift by restoring the collective Raman resonance
during the interval where the pump and Stokes pulses overlap most strongly.
The analytical condition in Eq.(\ref{eq:optimalchirp})  provides a direct prescription for selecting the chirp
rate required to achieve this compensation.

The numerical calculations demonstrate that applying the analytically
predicted chirp restores efficient population transfer while preserving the
same oscillatory dependence on the pulse area observed in the
noninteracting system.
This behavior confirms that the interaction does not alter the fundamental
collective transfer mechanism.
Instead, it merely shifts the location in parameter space at which that
mechanism becomes most effective.

The overall dynamics may therefore be viewed as the combined action of two
distinct physical processes.

First, the dressed-manifold structure determines how population is transported
through the collective Hilbert space.
This mechanism is intrinsic to the symmetric six-state system and is
responsible for the oscillatory dependence of the transfer efficiency on the
pulse area.

Second, the Rydberg--Rydberg interaction determines the instantaneous position
of the collective Raman resonance.
Frequency chirping compensates this interaction-induced detuning without
modifying the underlying dressed-manifold dynamics.

The separation of these two roles shows that collective chirped STIRAP can be understood as the interplay of
(i) coherent transport mediated by coupled dressed manifolds and
(ii) dynamic restoration of resonance through appropriately tailored chirped
laser fields.
This picture provides a unified framework for analyzing coherent population
transfer in interacting multilevel systems and is expected to remain
applicable to larger ensembles where multidimensional adiabatic manifolds
naturally arise.

%%%%%%%%%%%%%%%%%%%%%%%%%%%%%%%%%%%%%%%%%%%%%%%%%%%%%%%%%%%%%%%%%%%%%%%%%%%%%%
\section{Conclusions}

We have investigated collective chirped stimulated Raman adiabatic passage
(STIRAP) in a pair of identical three-level ladder atoms using a symmetric
six-state model and developed a comprehensive dressed-state interpretation of
the observed collective dynamics. The study was motivated by the unexpected
oscillatory dependence of the population transferred to the doubly excited
Rydberg state on the peak Rabi frequency, a behavior that has no direct
counterpart in conventional single-atom STIRAP.

%Our analysis demonstrates that this oscillatory transfer cannot be explained within a bare-state picture based on interference between alternative excitation pathways. Although the collective Hamiltonian permits several possible routes connecting the initial and target states, the corresponding bare-state populations remain strongly asymmetric during the evolution and do not support a simple pathway-interference interpretation.

Our analysis demonstrates that  population dynamics is governed by the instantaneous dressed-state
structure of the collective Hamiltonian. By introducing gauge-invariant
projectors onto the three instantaneous dressed manifolds, we have shown that
the wavefunction evolves predominantly within a two-dimensional dark manifold
while undergoing transient population exchange with a neighboring bright
manifold during the pulse-overlap interval. The outer dressed manifold
remains essentially inactive throughout the entire evolution.

To characterize this process quantitatively, we introduced a
wavefunction-projected nonadiabatic coupling between dressed manifolds,
providing a natural generalization of the familiar STIRAP mixing rate to
multidimensional adiabatic subspaces. The calculations reveal that this
coupling is strongly localized within the pulse-overlap region and is
responsible for the temporary communication between the dark and neighboring
bright manifolds. Following this transient exchange, the population returns
almost entirely to the dark manifold before the completion of the STIRAP
sequence, demonstrating that the bright manifold acts primarily as a mediator
of the collective evolution rather than as the final transport channel.

We further investigated the influence of the Rydberg--Rydberg interaction on
the collective transfer process. %The interaction shifts the energy of the doubly excited pair state and thereby modifies the instantaneous collective Raman resonance condition. 
From the effective Hamiltonian we derived a simple
analytical expression relating the optimal chirp rate to the interaction
strength and the desired resonance time. For the pulse sequence considered in
this work, the resulting condition provides a direct prescription for
selecting the chirp required to compensate the interaction-induced detuning.

Numerical simulations confirm that the analytically predicted chirp restores
efficient collective population transfer while preserving the oscillatory
dependence on the pulse area. These results demonstrate that the
Rydberg--Rydberg interaction does not alter the fundamental dressed-manifold
mechanism responsible for the oscillatory dynamics. Instead, it shifts the
location of the collective Raman resonance, whereas the chirp restores this
resonance without modifying the underlying manifold structure.

Viewed together, these results lead to a unified physical picture of collective chirped STIRAP. The coherent transport is governed by transient communication between neighboring dressed manifolds, while the role of the chirp is to dynamically compensate interaction-induced detuning and maintain the resonance conditions required for efficient transfer. The dressed-manifold framework developed here provides a physically transparent description of collective adiabatic dynamics and is expected to be applicable to larger interacting multilevel systems possessing multidimensional adiabatic subspaces.

%The present work also opens several directions for future investigation. Extension of the dressed-manifold formalism to larger atomic ensembles, inclusion of spatially dependent Rydberg interactions and blockade effects, and analysis of robustness in the presence of spontaneous emission and laser phase fluctuations represent natural next steps. More generally, the wavefunction-projected manifold approach introduced here may provide a useful framework for understanding coherent transport in a broad class of interacting quantum systems where conventional single-state adiabatic descriptions are no longer adequate.

Beyond the specific two-atom system considered here, these results suggest several potential applications of manifold-based adiabatic control. The ability to identify and selectively control transient communication between neighboring dressed manifolds may provide a useful strategy for preparing correlated Rydberg states and for designing collective excitation and entanglement protocols in systems where a single-dark-state description is no longer adequate. In this perspective, temporary population of a neighboring bright manifold need not be regarded solely as a breakdown of adiabatic following, but may instead serve as a controllable component of the transfer process. The chirp-compensation condition also suggests a complementary spectroscopic application: rather than using a known Rydberg--Rydberg interaction to determine the required chirp, the procedure may be inverted so that the chirp at which efficient collective transfer is restored provides a measure of the interaction-induced energy shift. % More generally, the gauge-invariant projector formulation developed here offers a natural route toward analyzing and designing coherent dynamics in larger interacting ensembles, where multidimensional and nearly degenerate dressed manifolds are expected to replace individual adiabatic eigenstates as the relevant dynamical objects.

% Future sections may be inserted below.
% \input{section_model}
% \input{section_collective_stirap}
% \input{section_dressed_manifold}
% \input{section_interaction}
% \input{section_conclusions}

\bibliographystyle{apsrev4-2}
\bibliography{AtomPairReferences}

\end{document}